\documentclass[aps,prl,twocolumn,preprintnumbers, superscriptaddress]{revtex4-2}
\pdfoutput=1

\usepackage[utf8]{inputenc}
\usepackage[T1]{fontenc}
\usepackage{amsmath,amssymb,amsthm,amsfonts,mathrsfs,mathtools}
\usepackage{ragged2e}
\usepackage{enumitem}
\usepackage{young,bm,braket}
\usepackage{dsfont}
\usepackage[vcentermath]{youngtab}
\usepackage{tikz}
\usetikzlibrary{shapes}
\usetikzlibrary{calc}
\usetikzlibrary{arrows.meta}
\usetikzlibrary{positioning}
\usetikzlibrary{decorations.markings}
\usetikzlibrary{decorations.pathmorphing}
\usetikzlibrary{fadings}

\allowdisplaybreaks

\usepackage{slashed}
\usepackage{hyperref}
\hypersetup{
	pdfstartview={FitH},    
	pdftitle={From Integrability to Chaos in Tensionless AdS$_3$ String Theory},    
	pdfauthor={Matthias R.~Gaberdiel, Beat Nairz, Cheng Peng},     
hidelinks}
\usepackage{graphicx}
\usepackage{tikz}

\usetikzlibrary{decorations.markings}
\tikzset{line/.style={line width=0.25mm},
	curve/.style={line,smooth,tension=1},
	->-/.style={decoration={
			markings,
			mark=at position #1 with {\arrow[>=stealth]{>}}},postaction={decorate}},
	-<-/.style={decoration={
			markings,
			mark=at position #1 with {\arrow[>=stealth]{<}}},postaction={decorate}},
}

\usepackage{times}
\usepackage[scaled]{couriers}

\usepackage{xcolor}
\usepackage{mdframed}

\renewenvironment{figure}[1][]{
	\begin{originalfigure}[#1]
		\begin{mdframed}[linecolor=black!0,backgroundcolor=black!0]
		}{
		\end{mdframed}
	\end{originalfigure}
}

\definecolor{MM1}{rgb}{0.368417, 0.506779, 0.709798}
\definecolor{MM2}{rgb}{0.880722, 0.611041, 0.142051}
\definecolor{MM3}{rgb}{0.560181, 0.691569, 0.194885}

\newcommand{\be}{\begin{equation}}
\newcommand{\ee}{\end{equation}}

\def\bal#1\eal{\begin{align}#1\end{align}}
\def\a{\alpha}

\usepackage{comment}

\DeclareMathAlphabet{\mathpzc}{OT1}{pzc}{m}{it}

\newcommand{\mrm}[1]{\mathrm{#1}}
\newcommand{\mcl}[1]{\mathcal{#1}}
\newcommand{\mbb}[1]{\mathbb{#1}}

\newcommand{\mfr}[1]{\mathfrak{#1}}

\newcommand{\cint}[1]{\underset{#1}{\oint}}

\newcommand{\br}[1]{\left({#1}\right)}
\newcommand{\bbr}[1]{\big({#1}\big)}
\newcommand{\corr}[1]{\left\langle{#1}\right\rangle}

\ExplSyntaxOn
\cs_new:Npn \pd {\partial}
\cs_new:Npn \ed {\mathrm d}
\cs_new:Npn \im#1 {\mathrm{im}\,{#1}}
\cs_new:Npn \Ric {\mathcal R\mathrm{ic}}
\cs_new:Npn \tr#1{
\str_if_eq:nnTF{#1}{R}
{\mathrm{tr}\c_math_subscript_token g\mathcal R\mathrm{ic}}
{\str_if_eq:nnTF{#1}{K}
{\mathrm{tr}\c_math_subscript_token h\mathcal K}
{\mathrm{tr}{#1}}
}
}
\cs_new:Npn \K {\mathscr K}
\cs_new:Npn \m {1,\cdots,m}
\cs_new:Npn \up#1#2{
\bool_if:nTF{\str_if_eq_p:nn{#1}{V} || \str_if_eq_p:nn{#1}{M}||\str_if_eq_p:nn{#1}{P}|| \str_if_eq_p:nn{#1}{O}}
{
{\mathscr #1}^{[#2]}
}
{
\bool_if:nTF{\str_if_eq_p:nn{#1}{K}||\str_if_eq_p:nn{#1}{R}||\str_if_eq_p:nn{#1}{G}||\str_if_eq_p:nn{#1}{T}}
{
{\mathcal #1}^{[#2]}
}
{
{#1}^{[#2]}
}
}
}
\cs_new:Npn \GN {G\c_math_subscript_token {\text N}}
\cs_new:Npn \diag {\mathrm{diag}}
\cs_new:Npn \Span {\mathrm{Span}}
\cs_new:Npn \SF {\mathsf{SF}}
\cs_new:Npn \II {\mathrm{I\!I}}
\cs_new:Npn \AdS {\text{AdS}}
\cs_new:Npn \ads#1 {\text{AdS}_#1}
\cs_new:Npn \Sch#1 {
\mathsf{Sch}\big[{#1}^{[i]}\big]
}
\ExplSyntaxOff

\begin{document}

\title{From Integrability to Chaos in Tensionless AdS3 String Theory}

\author{Matthias R.~Gaberdiel}
\affiliation{Institut f\"ur Theoretische Physik,
ETH Z\"urich,
Wolfgang-Pauli-Strasse 27,
8093 Z\"urich, Switzerland}
\affiliation{Kavli Institute for Theoretical Sciences (KITS) and School of Quantum, University of the Chinese Academy of Sciences, Beijing 100190, China}

\author{Beat Nairz}
\affiliation{Institut f\"ur Theoretische Physik,
ETH Z\"urich,
Wolfgang-Pauli-Strasse 27,
8093 Z\"urich, Switzerland}

\author{Cheng Peng}
\affiliation{Kavli Institute for Theoretical Sciences (KITS) and School of Quantum,
University of the Chinese Academy of Sciences, Beijing 100190, China}
\affiliation{Peng Huanwu Center for Fundamental Theory, Hefei, Anhui 230026, China}

\begin{abstract}
We study quantum-gravitational corrections to the planar integrable structure of the
symmetric product orbifold  $\text{Sym}^N(\mathbb{T}^4)$ that is dual to tensionless string theory on AdS$_3\times$S$^3\times \mathbb{T}^4$. We consider the deformation that switches on R-R flux in the bulk, and analyze the associated level spacing statistics. The leading non-planar corrections lift the planar degeneracies and produce signs of random-matrix behavior. Our results provide a controlled example of string interactions driving the onset of quantum chaos.
\end{abstract}

\maketitle

\section{Introduction}
A central question in holography is how chaotic gravitational dynamics emerge from microscopic quantum field theories. This question is especially sharp in models that are solvable at large $N$: the strict large-$N$ limit often comes with hidden symmetries and integrable structures, whereas dynamical gravity is expected to introduce graviton scattering, chaotic spectral statistics, topology changes, and black hole formation. A controlled framework capable of capturing the first steps away from integrability and the subsequent onset of chaos is therefore highly valuable.

The tensionless limit of string theory on $\text{AdS}_3\times {\rm S}^3\times \mbb{T}^4$ with one unit of NS-NS flux provides such a laboratory. It is dual to the symmetric product orbifold
$\text{Sym}^N\bbr{\mbb{T}^4}\equiv \left(\mathbb{T}^4\right)^N/S_N$~\cite{Gaberdiel:2018rqv,Eberhardt:2018ouy,Eberhardt:2019ywk,Dei:2020zui}, and the orbifold CFT has exactly marginal deformations preserving the $\mcl{N}=(4,4)$ superconformal symmetry. Moving away from the orbifold point along one of the moduli corresponds to turning on R-R flux in the bulk \cite{Gaberdiel:2025smz}. This deformation breaks the large stringy symmetry of the symmetric orbifold~\cite{Gaberdiel:2015uca,Keller:2019suk,Benjamin:2021zkn}, but in the planar limit a striking integrable structure remains \cite{Gaberdiel:2023lco,Gaberdiel:2024nge,Gaberdiel:2024dfw}.

Evidence for this planar integrability includes large degeneracies in the anomalous dimensions that are not explained by the $\mcl{N}=(4,4)$ superconformal algebra. Additionally, there are degeneracies in the limit of large twist studied in \cite{Gaberdiel:2023lco}, dual to the BMN limit \cite{Berenstein:2002jq}, that are of the same structural form as those appearing in integrable spin chain descriptions~\cite{Beisert:2003tq}. As we demonstrate in the following, the distinct planar eigenvalues exhibit Poisson statistics as expected for an integrable quantum system. We can therefore ask whether  these integrable structures survive once the bulk gravity, or its string theory completion, becomes dynamical. The strength of these quantum corrections is controlled by  $1/N$ in the dual field theory \footnote{The coupling constant $g_s$ of the dual AdS$_3$ string theory is proportional to $g_s\sim 1/\sqrt{N}$. However, the subleading anomalous conformal dimensions involve the splitting and joining of strings, see FIG.~\ref{fig:cov_surf}, and hence are suppressed by $g_s^2 \sim 1/N$.}; hence, these effects manifest themselves as non-planar $1/N$ corrections in the microscopic quantum field theory. Dynamically, these contributions arise from mixing with multi-cycle states in the symmetric orbifold, which provides the holographic dual to topology-changing interactions on the string worldsheet via higher-genus loops.

In this letter we study in detail the level spacing statistics of certain $\frac{1}{4}$-BPS states in the symmetric orbifold, and show that the non-planar corrections lead to the onset of random-matrix-like spectral statistics \footnote{In a companion paper \cite{PRDcomp} we have studied the behavior of the degeneracies to subleading order in $\frac{1}{N}$, and found direct evidence that integrability is indeed broken by these effects.}. These findings provide an analytically tractable, microscopic realization of gravitational corrections driving a system from planar integrability to quantum chaos. Furthermore, this transition manifests itself within a quantum and highly stringy regime: with only a single unit of NS-NS flux, the curvature radius of the background geometry is comparable to the string scale, and the $1/N$ corrections directly capture fundamental string loop effects.

\section{\texorpdfstring{$\text{Sym}^N\bbr{\mbb{T}^4}$}{SymN(T4)} and its marginal deformation}\label{sec:non-planar_corr}

The symmetric orbifold $\text{Sym}^N\bbr{\mbb{T}^4}$ is defined by $N$ copies of the seed $\mbb{T}^4$ theory whose fundamental fields include four left-moving bosons $\alpha^i,\bar{\alpha}^i$, $i=1,2$ and fermions $\psi^\pm,\bar{\psi}^\pm$, together with the right-movers distinguished by a tilde. The fields satisfy the OPE relations
\bal
\nonumber \bar{\alpha}^i(x)\alpha^j(y)\sim \frac{\epsilon^{ij}}{\big(x-y\big)^2}\ ,\quad \bar{\psi}^\pm(x)\psi^\mp(y)\sim \frac{\pm 1}{x-y} \ .
\eal
The (left-moving) $\mcl{N}=4$ supercharges are defined as
\bal
\nonumber & G^+ =  \bar{\alpha}^2\,\psi^++\alpha^2\,\bar{\psi}^+\ , \quad
G'^+ = -\bar{\alpha}^1\,\psi^+-\alpha^1\,\bar{\psi}^+\ ,\\
\nonumber & G^- = \bar{\alpha}^1\,\psi^-+\alpha^1\,\bar{\psi}^-\ , \quad
G'^- = \bar{\alpha}^2\,\psi^-+\alpha^2\,\bar{\psi}^-\ .
\eal
Due to the $S_N$ quotient, there are twisted sectors whose fields have non-trivial monodromy; in a $w$-cycle twisted sector, the fundamental fields are fractionally moded, e.g. $\a^i_{{n}/{w}}$, $n\in \mathbb{Z}$. We shall study $S_N$-invariant excitations built on the BPS ground state $\ket{w}$ in this twisted sector, which has left- and right-moving conformal dimension and $\mathfrak{su}(2)$ charges all equal $h=j=\frac{w-1}{2}=\tilde{h}=\tilde{j}$.

The theory admits an exactly marginal deformation
$\Delta S= g\int d^2 x\,\Phi(x,\bar{x})$
~\cite{Dixon:1987bg}, where $g$ is the coupling constant and
\bal
\ket{\Phi} =
\frac{i}{\sqrt{2}}\,\bigl(G^-_{-\frac{1}{2}}\tilde{G}'^-_{-\frac{1}{2}}-
G'^-_{-\frac{1}{2}}\tilde{G}^-_{-\frac{1}{2}}
\bigr)\ket{2}\ ,\label{pert}
\eal
with $\ket{2}$ denoting the BPS ground state in the 2-cycle twisted sector. This deformation is dual to switching on R-R flux in the string background.

\subsection{Anomalous dimensions in the sector descending from $1/4$-BPS states}

We compute anomalous dimensions of single-particle states $\ket{\phi}$ which are purely left-moving descendants of the BPS state $\ket{w}$ in the $w$-cycle twisted sector. At the symmetric orbifold point these states are $1/4$-BPS, but after the marginal deformation the multiplets can recombine and their dimensions need not remain protected. The anomalous dimension matrix, namely the difference between the conformal dimension and the $\mathfrak{su}(2)$ charge
\begin{equation}
\tilde{\gamma} = \tilde{L}_0 - \tilde{K}^3_0= \{\tilde{G}'^+_{-\frac{1}{2}},\tilde{G}'^-_{\frac{1}{2}}\} \ ,\label{gt}
\end{equation}
can be expanded at large $N$ as
\begin{equation}\label{eq:gamma_expansion}
\tilde{\gamma} = \frac{1}{N}\,\tilde{\gamma}_0 + \frac{1}{N^2}\br{\tilde{\gamma}_1 + d_0\,\tilde{\gamma}_0} + \dots\ ,
\end{equation}
{where $\tilde{\gamma}_0$ is the planar anomalous dimension matrix computed at order $g^2$ in the coupling as in~\cite{Gaberdiel:2023lco}, $\tilde{\gamma}_1$ is the leading non-planar correction at $\mcl{O}\bbr{g^2}$}, and $d_0$ is an expansion coefficient from the planar normalization constant \cite{Pakman:2009zz}~{\footnote{Note that the expansions in $g$ and $1/N$ have distinct origins:  
$g$ is the coupling constant of the perturbation by $\ket{\Phi}$, and to lowest non-trivial order, the anomalous conformal dimensions appear at second order in the perturbation, i.e.\ are proportional to $g^2$. On the other hand, the factors of $N$ come from the combinatorics of the contractions and distinguish, in particular, single-cycle and double-cycle intermediate states.}}.
We are interested in the perturbative $1/N$ expansion of anomalous dimensions, i.e.~the eigenvalues of $\tilde{\gamma}$,
\begin{equation}\label{eq:anomalous_dimension_expansion}
\epsilon = \frac{1}{N}\,\epsilon_0 + \frac{1}{N^2}\,\bbr{\epsilon_1+d_0\,\epsilon_0} + \cdots \ .
\end{equation}
The planar anomalous dimension $\epsilon_0$ is obtained by diagonalizing $\tilde{\gamma}_0$ using the methods of \cite{Gaberdiel:2024nge}. The perturbative corrections $\epsilon_1$ can then be found by diagonalizing $\tilde{\gamma}_1$ on the eigenspaces of $\tilde{\gamma}_0$.
The techniques to compute the relevant $1/N$ corrections are developed in the companion paper~\cite{PRDcomp}, where the degeneracies of certain $1/4$-BPS states were studied in detail. We outline the method in the appendix.

As we will show below, the statistics of $\tilde{\gamma}_0$ reveal the planar integrable structure, while the statistics of the non-planar correction $\tilde{\gamma}_1$ show level repulsion characteristic of chaotic spectra. For definiteness, we consider $\tilde{\gamma}$ on two single-particle subspaces $\mbb{V}_\mrm{b}$, $\mbb{V}_\mrm{f}$ containing excitations of the BPS vacuum in the $w$-twisted sector consisting of two bosonic and fermionic magnons, respectively,
\begin{align}\label{eq:reference_states}
\mbb{V}_\mrm{b} &\ni \alpha^1_{-\frac{m}{w}}\alpha^1_{-1+\frac{m}{w}}\ket{w}\ , & & & \nonumber\\
\mbb{V}_\mrm{f} &\ni  \psi^-_{\frac{1}{2}-\frac{m}{w}}\psi^-_{-\frac{1}{2}+\frac{m}{w}}\ket{w}\ , & m&=1,\dots,w-1\ .\end{align}
{We will refer to these families of states as ``bosonic'' and ``fermionic'' to distinguish their different constituent magnons; however, both are bosonic.   The two families are not related by the global $\mcl{N}=(4,4)$ algebra of the CFT, and thus their anomalous dimensions need not coincide.}\\
In this letter, we will focus on the level statistics of these states, and show that they exhibit a transition from Poisson statistics to random matrix behavior as the $1/N$ corrections are included. We will also find strong evidence that the integrable structure in the planar limit is even present at finite twist $w$, something that
could not be analyzed by the methods of \cite{PRDcomp}. We are in principle able to carry out the calculation at any $w$, but due to the fast growth of the number of mixings, we are in practice limited to $w\leq 11$.

\subsection{\texorpdfstring{$1/N$}{1/N} expansion and multi-cycle states}

\begin{figure}[ht]
\centering
\includegraphics[width=.9\textwidth]{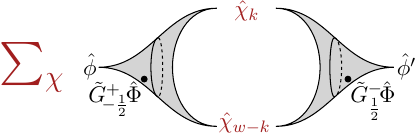}
\caption{A sketch of the covering surface representing the order-$g^2$ nonplanar contribution to the anomalous-dimension matrix $\bra{\phi'}\tilde{\gamma}\ket{\phi}$ under the perturbation $\Phi$. The non-planar corrections in Eq.~(\ref{eq:gamma_decomposition}) can be calculated by summing over all transitions where the intermediate states lie in a multi-cycle twisted sector.
As the covering surface can be identified with the string worldsheet \cite{Eberhardt:2019ywk}, this corresponds to string splitting, capturing a stringy process with  non-trivial worldsheet topology.}\label{fig:cov_surf}
\end{figure}

The anomalous dimension matrix element of Eq.~(\ref{gt}) between two states $\ket{\phi} $ and $\ket{\phi'}$ can be computed as
\begin{align}
\nonumber \bra{\phi'}\tilde{\gamma}\ket{\phi} =& \sum_{\chi} \bra{\phi'}\tilde{G}'^+_{-\frac{1}{2}}\ket{\chi}_{\Phi} \bra{\chi}\tilde{G}'^-_{\frac{1}{2}}\ket{\phi}_{\Phi} \\
&\qquad + \bra{\phi'}\tilde{G}'^-_{\frac{1}{2}}\ket{\chi}_{\Phi}\bra{\chi}\tilde{G}'^+_{-\frac{1}{2}}\ket{\phi}_{\Phi}\ ,\label{eq:gamma_decomposition}
\end{align}
where ${(\cdot)}_{\Phi}$ denotes the first-order perturbed three-point function, see e.g.~\cite[Eq.~(3.2)]{Gaberdiel:2024nge}, and one sums over a complete basis $\{\chi\}$. The $1/N$ expansion (\ref{eq:gamma_expansion}) of $\tilde{\gamma}$ is controlled by the cycle structure of intermediate states $\chi$ \cite{Pakman:2009zz}. The planar contribution is composed of single-cycle intermediate states, while the leading $1/N$ order correction comes from multi-cycle twisted sector states $\chi$. Concretely, these states arise from the action of the 2-cycle perturbing field $\Phi$ on two non-adjacent colors of the $w$-cycle state $\ket{\phi}$ leading to a disjoint $k$ and $w-k$ cycle,
\begin{equation}
(1\,k+1)\cdot (1\cdots w) = (1\cdots k)(k+1\cdots w)\ ,
\end{equation}
where $k=2,\dots, w-2$.
Given this structure, the intermediate states $\chi$ in the sum of~\eqref{eq:gamma_decomposition} that contribute to the $\tilde{\gamma}_1$ correction are magnon excitations on the multi-cycle BPS ground states
\bal
\ket{w-k,k}\equiv \ket{w-k}\otimes\ket{k}\ ,
\eal
and the constituent three-point functions can be computed using the covering map method~\cite{Lunin:2000yv}. The covering surface is sketched in Figure~\ref{fig:cov_surf}. In bulk language, this is the first place where string splitting/joining processes enter the anomalous-dimension computation.\\
In the appendix, we explain the essential steps in calculating the matrix elements $\bra{\phi'}\tilde{\gamma}\ket{\phi}$ using the covering map method.

\begin{figure*}[ht]
\centering
\begin{minipage}[t]{.44\textwidth}
\begin{tikzpicture}
\node[inner sep=0pt](n1)at(0,0){\includegraphics[width=\textwidth]{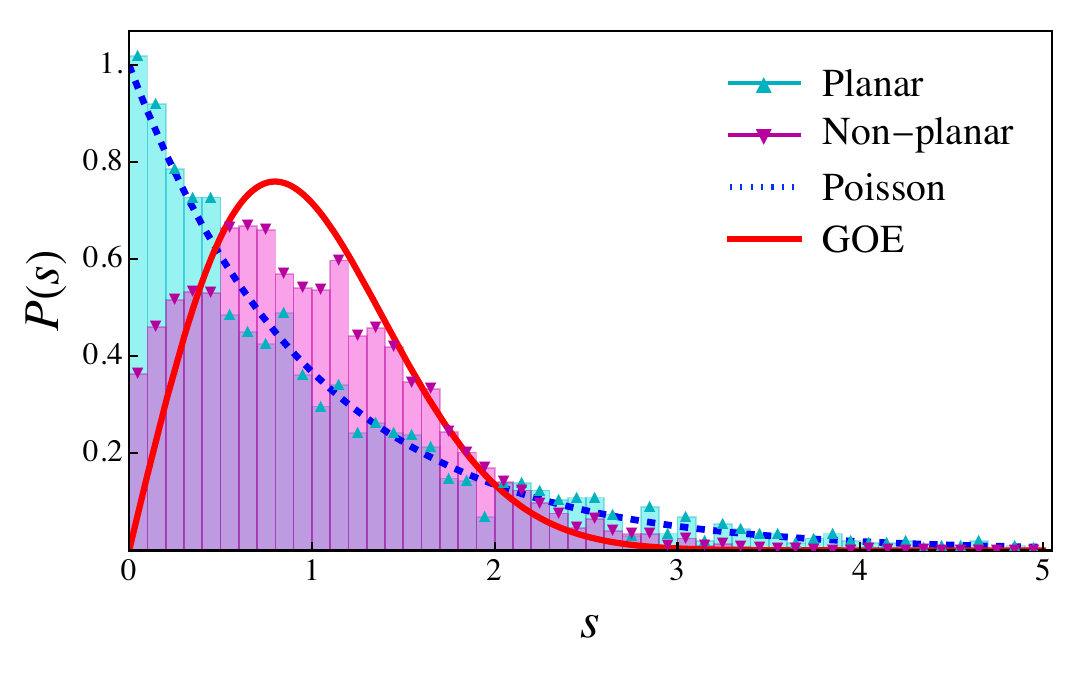}};
\node[anchor=north west]at(n1.north west){\large(a)};
\node[rotate=90,fill=white]at([xshift=0.9em,yshift=1em]n1.west){\large$P_\text{b}(s)$};
\end{tikzpicture}
\end{minipage}
\hfill
\begin{minipage}[t]{.44\textwidth}
\begin{tikzpicture}
\node[inner sep=0pt](n1)at(0,0){\includegraphics[width=\textwidth]{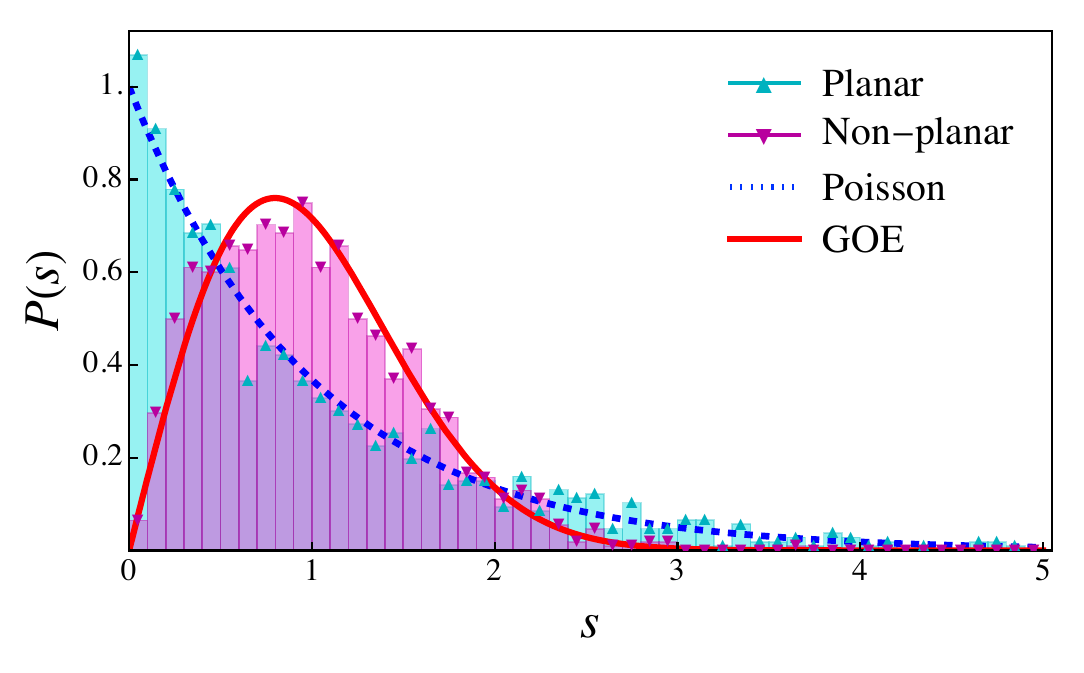}};
\node[anchor=north west]at(n1.north west){\large(b)};
\node[rotate=90,fill=white]at([xshift=0.9em,yshift=1em]n1.west){\large$P_\text{f}(s)$};
\end{tikzpicture}
\end{minipage}

\vspace*{-1em}
\caption{The probability distribution $P(s)$ of the unfolded level spacing $s$ for the planar and non-planar spectra of bosonic states $\mbb{V}_\mrm{b}$ in (a) and fermionic states $\mbb{V}_\mrm{f}$ in (b) at $w=11$. The cyan histogram (with upward-pointing triangles) describes the planar level spacing distribution, while the magenta histogram (with downward-pointing triangles) shows that of the non-planar corrections. For comparison, we overlay a Poisson distribution and the Wigner surmise of the GOE. The planar spectra are well described by Poisson statistics in both cases, as expected for an integrable system. For the bosons (a), the combined statistics of level spacings of the non-planar corrections deviate from Poisson and move toward level repulsion, but remain in an intermediate crossover regime. The non-planar corrections of the fermions (b) show stronger level repulsion and are close to the GOE Wigner surmise.}\label{fig:levelspacing_bosons}
\end{figure*}

\section{Planar degeneracies at finite $w$}\label{sec:pladeg}

In the following we shall concentrate on the statistical properties of the spectrum of the bosonic and fermionic states in Eq.~(\ref{eq:reference_states}). We first compute the planar anomalous dimension spectrum $\text{spec}\br{\tilde{\gamma}_0}$ of all such states explicitly at finite twist length $w$.

A striking feature of the planar anomalous dimensions is the large degeneracy of the spectrum, even at finite  twist length $w$. For example, at $w=11$, the dimension of $\mbb{V}_\mrm{b}$ in Eq.~(\ref{eq:reference_states}) is $10{,}472$, but there are only $2{,}034$ different anomalous dimension eigenvalues $\epsilon_0$. The fermionic sector contains $5{,}498$ states but only $1{,}074$ distinct eigenvalues. These systematic degeneracies provide evidence of a large residual symmetry and an underlying integrable structure, even at finite $w$.

\section{Level Spacing Statistics}\label{sec:chaos}

The non-planar corrections $\tilde{\gamma}_1$ lift the degeneracies among all of these states and exhibit level repulsion; they are thus reminiscent of random matrix behavior.
It is therefore interesting to analyse the precise statistical properties of these non-planar corrected eigenvalues and check whether they manifest explicit signatures of quantum chaos.

Indeed, tracking fine-grained spectral statistics, such as nearest-neighbor level spacings, serves as a universal diagnostic for chaotic behavior. While integrable systems typically exhibit Poisson statistics, chaotic quantum systems possess spectral correlations governed by random matrix theory (RMT)~\cite{Haake:2010fgh}. In the context of holography, RMT statistics are closely intertwined with the spectral fluctuations of black holes~\cite{Sekino:2008he, Shenker:2013pqa}, as recently exemplified by Jackiw--Teitelboim gravity and its holographic dual, the Sachdev--Ye--Kitaev model~\cite{kitaev2015simple, Maldacena:2016hyu, You:2016ldz, Garcia-Garcia:2016mno, Cotler:2016fpe, Stanford:2019vob, Witten:2020wvy, Turiaci:2023jfa}. A related crossover has also been observed in $\mathcal{N}=4$ super-Yang--Mills at finite $N$, where the level spacing statistics of the $\mathfrak{sl}(2)$ and $\mathfrak{su}(2)$ sectors (evaluated at the one- and two-loop levels) exhibit the Gaussian Orthogonal Ensemble (GOE) level spacing statistics~\cite{McLoughlin:2020zew}. {The lifting of the planar degeneracies in the $\mathfrak{su}(2)\times
\mathfrak{su}(2)$ sector of ABJM was also studied in \cite{Kristjansen:2008ABJM}; however, the spectral statistics was not analysed there.}

In practice, we analyse the level spacing of the non-planar corrections around a highly degenerate eigenspace $\mcl{E}$ consisting of all states sharing a common planar anomalous dimension $\epsilon_0$. The non-planar corrections $\epsilon_1$ are obtained by diagonalising the non-planar correction matrix  $\tilde{\gamma}_1\big|_{\mcl{E}}$.
In order to compare this spectrum to RMT, we unfold the spectrum following~\cite{Guhr:1997ve} and denote the unfolded eigenvalues by $E_i$. The unfolded level spacing is the difference between successive ordered eigenvalues
$s_i=E_{i+1}-E_{i}$.
The probability distribution of the level spacings on $\mcl{E}$ is denoted by $P_{\mcl{E}}(s)$. We compute this distribution for all sufficiently large degenerate eigenspaces ($\text{dim}(\mcl{E})>10$)---smaller dimensional eigenspaces lead to large statistical fluctuations---and then combine the level spacing distributions into a single histogram
of $5{,}491$ bosonic spacings and $1{,}129$ fermionic spacings, to obtain combined statistics. We comment on the technical details about the unfolding procedure in the appendix.

We further compute another statistical quantity, namely the gap ratios
$r_i = \frac{\min(s_i,s_{i+1})} {\max(s_i,s_{i+1})}$, that can be used to diagnose the overall shape of the level spacing distribution~\footnote{For spectra with slowly varying average density, the gap ratio is expected to be independent of unfolding. We have checked that $\protect\langle r \protect\rangle$ varies by less than $10^{-4}$ when using raw or unfolded eigenvalues, suggesting that the unfolding procedure is consistent.}. The average gap ratios for Poisson statistics and the Gaussian orthogonal ensemble in RMT are~\cite{Atas:2013gvn}
\bal
\langle r\rangle_{\text{Poisson}}= 0.386\ ,\quad \langle r\rangle_{\text{GOE}} = 0.531\ .
\eal

In Figure~\ref{fig:levelspacing_bosons}, we show the level spacing $P(s)$ distributions for the (distinct) planar anomalous dimensions and the non-planar corrections.
The planar eigenvalues are well-described by a Poisson distribution, consistent with integrability \cite{Haake:2010fgh}. The non-planar corrections deviate from Poisson and show signs of level repulsion. For the bosonic states $\mbb{V}_\text{b}$, the repulsion is not yet fully GOE-like. The average gap ratios for the planar and non-planar distributions (for the bosons) are
\bal
\nonumber \langle r\rangle_{\rm b,p}=0.388(7)\ ,\quad \langle r\rangle_{\rm b,np}=0.448(7)\ .
\eal
where we include bootstrap standard deviations which give an uncertainty estimate of the average~\footnote{For the planar eigenvalues, we implement a block bootstrap, realised by partitioning the gap ratios into blocks from which one randomly chooses with replacement. For the non-planar data,
we instead implement a cluster bootstrap, by randomly drawing
from the smaller $\tilde{\gamma}_1|_{\mcl{E}}$ matrices.}.
We therefore interpret the bosonic data as evidence for a crossover regime: the spectrum has moved away from the planar Poisson value, but it has not reached fully developed GOE statistics at $w=11$~\footnote{There have been previous observations of transitions between chaotic and integrable behavior in many examples of holographic disordered models \cite{Peng:2018zap,Chang:2021fmd,Chang:2021wbx,Berkooz:2022dfr, Berkooz:2024evs, Gao:2024lve}. These obervations suggest that large symmetries or integrable structures can emerge at special points in the moduli space of chaotic theories. Interestingly, non-maximal chaotic behavior was found for certain 2D and 3D disordered models in~\cite{Peng:2018zap,Chang:2021wbx}.}.

We also plot the result for a similar analysis of the fermionic states in $\mbb{V}_\text{f}$. The planar level spacings are again consistent with a Poisson distribution. The non-planar level spacings show clearer RMT behavior, with stronger level repulsion, and are close to the GOE distribution. The average gap ratios, within the accessible finite-size data, are
\begin{equation}
\nonumber \langle r\rangle_{\rm f,p}=0.362(9)\ ,\quad \langle r\rangle_{\rm f,np}=0.523(9)\ .
\end{equation}
The fermionic sector therefore exhibits the onset of random-matrix statistics more strongly. In the appendix, we present further statistical evidence for our findings by describing the gap ratio distributions and carrying out a Cram\'er--von Mises test. We should mention that the appearance of a GOE distribution is natural in view of the fact that the perturbing field is time-reversal invariant~\cite{Haake:2010fgh}~\footnote{Notice that the appearance of chaotic behavior in some BPS sector could be related to recent studies of BPS chaos \cite{Chen:2024oqv}, diagnosed by the LMRS prescription \cite{Lin:2022rzw,Lin:2022zxd} or recently also by a Berry curvature \cite{Chen:2026vml}, and the fortuity phenomena \cite{Chang:2013fba, Chang:2022mjp, Chang:2024lxt} especially in the orbifold CFT setup~\cite{Chang:2025rqy,Chang:2025wgo}.}.

\section{Discussion}

In this letter, we have computed the leading non-planar corrections to the anomalous dimension matrix for two families of states that are $1/4$-BPS at the symmetric orbifold point. Our findings are twofold. First, the non-planar terms systematically lift the degeneracies of the planar spectrum that are not a consequence of the superconformal symmetry. Second, the resulting spectra of non-planar splittings exhibit clear level repulsion and a pronounced departure from Poisson statistics; more specifically, the fermionic sector we have studied lies in close proximity to universal GOE statistics, while the corresponding bosonic sector occupies an intermediate crossover regime.

The physical picture that emerges is that the planar, large-$N$ theory retains an integrable structure even under the R-R flux deformation, whereas the first genuinely gravitational corrections break this structure. Microscopically within the CFT, these corrections are mediated by transitions through multi-cycle intermediate states in~\eqref{eq:gamma_decomposition}; holographically in the bulk, they map to string splitting and joining interactions controlled by $G_N \sim 1/N$. Crucially, this mechanism indicates that the lifting of degeneracies is not driven by a collection of arbitrary, statistically independent perturbations, but is instead dictated in a precisely calculable manner by the $1/N$ gravitational interactions of the holographic theory controlled by a single parameter. 
{The $\text{AdS}_3/\text{CFT}_2$ duality therefore provides a rare,  controlled microscopic window into how a tensionless holographic system begins its transition from integrability toward quantum chaos once $1/N$ gravitational interactions are turned on. Furthermore, our  tractable setup differs in interesting and fundamental ways from similar investigations in AdS/CFT. First, the integrable structure governing the planar limit of the deformed symmetric orbifold has not been understood from a spin-chain perspective, and therefore seems to have a somewhat different nature as that of $\mcl{N}=4$ SYM or ABJM, see \cite{Beisert:2010jr} for an overview~{\footnote{In particular, the symmetric orbifold theory seems to be vector-like (rather than a gauge-theory) in that the central charge is proportional to $N$, rather than $N^2$.}}. Second, the chaotic properties of $\mcl{N}=4$ SYM~\cite{McLoughlin:2020zew} were observed directly at finite $N$, whereas in our analysis we find chaotic signatures as soon as the leading perturbative $1/N$ corrections
are included; in particular, our results do not rely on the finite-$N$ stringy exclusion effects of \cite{Maldacena:1998bw}. 
Finally, the detailed duality of \cite{Gaberdiel:2018rqv, Eberhardt:2018ouy,Eberhardt:2019ywk} allows us to interpret our findings directly in a theory of quantum gravity. Our work is thus a valuable prototype for understanding how gravitational interactions manifest themselves as chaotic dynamics across other holographic frameworks.}

Naturally, certain limitations apply to our current approach. The calculation is perturbative in $1/N$ and is valid in the regime $w \ll N$. Furthermore, explicit numerical computations have so far only been carried out for low twist, up to $w=11$, so the states studied here lie far below the black hole threshold; the present data should therefore be interpreted as a microscopic signature for the onset of chaotic spectral correlations rather than full thermalization. On the other hand, the sectors we have analyzed also have certain advantages: quantum fluctuations need not compete with a large classical energy to exhibit structural information extending beyond generic random-matrix universality \footnote{An illustrative example of this paradigm is the Sachdev-Ye-Kitaev model~\cite{Sachdev:1992fk, kitaev2015simple}, where the light states near the spectral edge carry additional structure of the Schwarzian sector that dominates at low temperatures and maps directly to Jackiw-Teitelboim gravity~\cite{Maldacena:2016hyu,Maldacena:2016upp, Garcia-Garcia:2016mno}. }. This then provides an efficient avenue to directly study the emergence of chaotic behavior from integrability that is driven entirely by quantum effects.

\noindent {\bf Note added:} As this paper was being finalized, we were informed by Haoyu Zhang that he had obtained similar results by studying the perturbed low-lying quarter-BPS spectrum for small values of $N$ \cite{Zhang:2026jnf}.

\section*{Acknowledgments} We thank Yiming Chen, Frank Coronado, Bin Guo, Ji Hoon Lee, Anthony Houppe,  Juan Maldacena, Edward Mazenc, Kiarash Naderi, Vit Sriprachyakul, and Haoyu Zhang for useful conversations, and Haoyu Zhang for helpful comments on the draft. MRG thanks the Institute for Advanced Study, Princeton, for hospitality during the final stages of this work; his stay there was supported by the Adler Family Fund. The work of BN is supported through a personal grant awarded to MRG from the Swiss National Science Foundation. The work of the group at ETH is also supported in part by the Simons Foundation grant 994306 (Simons Collaboration on Confinement and QCD Strings), as well as the NCCR SwissMAP that is also funded by the Swiss National Science Foundation. CP is supported by NSFC Nos.~12675092,  12175237, 12447108, 12247103 and funds from the Chinese Academy of Sciences.

\bibliographystyle{apsrev4-2}
\bibliography{PRL}

\begin{thebibliography}{61}%
\makeatletter
\providecommand \@ifxundefined [1]{%
 \@ifx{#1\undefined}
}%
\providecommand \@ifnum [1]{%
 \ifnum #1\expandafter \@firstoftwo
 \else \expandafter \@secondoftwo
 \fi
}%
\providecommand \@ifx [1]{%
 \ifx #1\expandafter \@firstoftwo
 \else \expandafter \@secondoftwo
 \fi
}%
\providecommand \natexlab [1]{#1}%
\providecommand \enquote  [1]{``#1''}%
\providecommand \bibnamefont  [1]{#1}%
\providecommand \bibfnamefont [1]{#1}%
\providecommand \citenamefont [1]{#1}%
\providecommand \href@noop [0]{\@secondoftwo}%
\providecommand \href [0]{\begingroup \@sanitize@url \@href}%
\providecommand \@href[1]{\@@startlink{#1}\@@href}%
\providecommand \@@href[1]{\endgroup#1\@@endlink}%
\providecommand \@sanitize@url [0]{\catcode `\\12\catcode `\$12\catcode
  `\&12\catcode `\#12\catcode `\^12\catcode `\_12\catcode `\%12\relax}%
\providecommand \@@startlink[1]{}%
\providecommand \@@endlink[0]{}%
\providecommand \url  [0]{\begingroup\@sanitize@url \@url }%
\providecommand \@url [1]{\endgroup\@href {#1}{\urlprefix }}%
\providecommand \urlprefix  [0]{URL }%
\providecommand \Eprint [0]{\href }%
\providecommand \doibase [0]{https://doi.org/}%
\providecommand \selectlanguage [0]{\@gobble}%
\providecommand \bibinfo  [0]{\@secondoftwo}%
\providecommand \bibfield  [0]{\@secondoftwo}%
\providecommand \translation [1]{[#1]}%
\providecommand \BibitemOpen [0]{}%
\providecommand \bibitemStop [0]{}%
\providecommand \bibitemNoStop [0]{.\EOS\space}%
\providecommand \EOS [0]{\spacefactor3000\relax}%
\providecommand \BibitemShut  [1]{\csname bibitem#1\endcsname}%
\let\auto@bib@innerbib\@empty
\bibitem [{\citenamefont {Gaberdiel}\ and\ \citenamefont
  {Gopakumar}(2018)}]{Gaberdiel:2018rqv}%
  \BibitemOpen
  \bibfield  {author} {\bibinfo {author} {\bibfnamefont {M.~R.}\ \bibnamefont
  {Gaberdiel}}\ and\ \bibinfo {author} {\bibfnamefont {R.}~\bibnamefont
  {Gopakumar}},\ }\href {https://doi.org/10.1007/JHEP05(2018)085} {\bibfield
  {journal} {\bibinfo  {journal} {JHEP}\ }\textbf {\bibinfo {volume} {05}},\
  \bibinfo {pages} {085}},\ \Eprint {https://arxiv.org/abs/1803.04423}
  {arXiv:1803.04423 [hep-th]} \BibitemShut {NoStop}%
\bibitem [{\citenamefont {Eberhardt}\ \emph {et~al.}(2019)\citenamefont
  {Eberhardt}, \citenamefont {Gaberdiel},\ and\ \citenamefont
  {Gopakumar}}]{Eberhardt:2018ouy}%
  \BibitemOpen
  \bibfield  {author} {\bibinfo {author} {\bibfnamefont {L.}~\bibnamefont
  {Eberhardt}}, \bibinfo {author} {\bibfnamefont {M.~R.}\ \bibnamefont
  {Gaberdiel}},\ and\ \bibinfo {author} {\bibfnamefont {R.}~\bibnamefont
  {Gopakumar}},\ }\href {https://doi.org/10.1007/JHEP04(2019)103} {\bibfield
  {journal} {\bibinfo  {journal} {JHEP}\ }\textbf {\bibinfo {volume} {04}},\
  \bibinfo {pages} {103}},\ \Eprint {https://arxiv.org/abs/1812.01007}
  {arXiv:1812.01007 [hep-th]} \BibitemShut {NoStop}%
\bibitem [{\citenamefont {Eberhardt}\ \emph {et~al.}(2020)\citenamefont
  {Eberhardt}, \citenamefont {Gaberdiel},\ and\ \citenamefont
  {Gopakumar}}]{Eberhardt:2019ywk}%
  \BibitemOpen
  \bibfield  {author} {\bibinfo {author} {\bibfnamefont {L.}~\bibnamefont
  {Eberhardt}}, \bibinfo {author} {\bibfnamefont {M.~R.}\ \bibnamefont
  {Gaberdiel}},\ and\ \bibinfo {author} {\bibfnamefont {R.}~\bibnamefont
  {Gopakumar}},\ }\href {https://doi.org/10.1007/JHEP02(2020)136} {\bibfield
  {journal} {\bibinfo  {journal} {JHEP}\ }\textbf {\bibinfo {volume} {02}},\
  \bibinfo {pages} {136}},\ \Eprint {https://arxiv.org/abs/1911.00378}
  {arXiv:1911.00378 [hep-th]} \BibitemShut {NoStop}%
\bibitem [{\citenamefont {Dei}\ \emph {et~al.}(2021)\citenamefont {Dei},
  \citenamefont {Gaberdiel}, \citenamefont {Gopakumar},\ and\ \citenamefont
  {Knighton}}]{Dei:2020zui}%
  \BibitemOpen
  \bibfield  {author} {\bibinfo {author} {\bibfnamefont {A.}~\bibnamefont
  {Dei}}, \bibinfo {author} {\bibfnamefont {M.~R.}\ \bibnamefont {Gaberdiel}},
  \bibinfo {author} {\bibfnamefont {R.}~\bibnamefont {Gopakumar}},\ and\
  \bibinfo {author} {\bibfnamefont {B.}~\bibnamefont {Knighton}},\ }\href
  {https://doi.org/10.1007/JHEP02(2021)081} {\bibfield  {journal} {\bibinfo
  {journal} {JHEP}\ }\textbf {\bibinfo {volume} {02}},\ \bibinfo {pages}
  {081}},\ \Eprint {https://arxiv.org/abs/2009.11306} {arXiv:2009.11306
  [hep-th]} \BibitemShut {NoStop}%
\bibitem [{\citenamefont {Gaberdiel}\ and\ \citenamefont
  {Meur}(2026)}]{Gaberdiel:2025smz}%
  \BibitemOpen
  \bibfield  {author} {\bibinfo {author} {\bibfnamefont {M.~R.}\ \bibnamefont
  {Gaberdiel}}\ and\ \bibinfo {author} {\bibfnamefont {I.~L.}\ \bibnamefont
  {Meur}},\ }\href {https://doi.org/10.1007/JHEP03(2026)231} {\bibfield
  {journal} {\bibinfo  {journal} {JHEP}\ }\textbf {\bibinfo {volume} {03}},\
  \bibinfo {pages} {231}},\ \Eprint {https://arxiv.org/abs/2509.03132}
  {arXiv:2509.03132 [hep-th]} \BibitemShut {NoStop}%
\bibitem [{\citenamefont {Gaberdiel}\ \emph {et~al.}(2015)\citenamefont
  {Gaberdiel}, \citenamefont {Peng},\ and\ \citenamefont
  {Zadeh}}]{Gaberdiel:2015uca}%
  \BibitemOpen
  \bibfield  {author} {\bibinfo {author} {\bibfnamefont {M.~R.}\ \bibnamefont
  {Gaberdiel}}, \bibinfo {author} {\bibfnamefont {C.}~\bibnamefont {Peng}},\
  and\ \bibinfo {author} {\bibfnamefont {I.~G.}\ \bibnamefont {Zadeh}},\ }\href
  {https://doi.org/10.1007/JHEP10(2015)101} {\bibfield  {journal} {\bibinfo
  {journal} {JHEP}\ }\textbf {\bibinfo {volume} {10}},\ \bibinfo {pages}
  {101}},\ \Eprint {https://arxiv.org/abs/1506.02045} {arXiv:1506.02045
  [hep-th]} \BibitemShut {NoStop}%
\bibitem [{\citenamefont {Keller}\ and\ \citenamefont
  {Zadeh}(2020)}]{Keller:2019suk}%
  \BibitemOpen
  \bibfield  {author} {\bibinfo {author} {\bibfnamefont {C.~A.}\ \bibnamefont
  {Keller}}\ and\ \bibinfo {author} {\bibfnamefont {I.~G.}\ \bibnamefont
  {Zadeh}},\ }\href {https://doi.org/10.1007/s00220-020-03721-4} {\bibfield
  {journal} {\bibinfo  {journal} {Commun. Math. Phys.}\ }\textbf {\bibinfo
  {volume} {377}},\ \bibinfo {pages} {225} (\bibinfo {year} {2020})},\ \Eprint
  {https://arxiv.org/abs/1905.00035} {arXiv:1905.00035 [hep-th]} \BibitemShut
  {NoStop}%
\bibitem [{\citenamefont {Benjamin}\ \emph {et~al.}(2021)\citenamefont
  {Benjamin}, \citenamefont {Keller},\ and\ \citenamefont
  {Zadeh}}]{Benjamin:2021zkn}%
  \BibitemOpen
  \bibfield  {author} {\bibinfo {author} {\bibfnamefont {N.}~\bibnamefont
  {Benjamin}}, \bibinfo {author} {\bibfnamefont {C.~A.}\ \bibnamefont
  {Keller}},\ and\ \bibinfo {author} {\bibfnamefont {I.~G.}\ \bibnamefont
  {Zadeh}},\ }\href {https://doi.org/10.1007/JHEP10(2021)089} {\bibfield
  {journal} {\bibinfo  {journal} {JHEP}\ }\textbf {\bibinfo {volume} {10}},\
  \bibinfo {pages} {089}},\ \Eprint {https://arxiv.org/abs/2107.00655}
  {arXiv:2107.00655 [hep-th]} \BibitemShut {NoStop}%
\bibitem [{\citenamefont {Gaberdiel}\ \emph {et~al.}(2024)\citenamefont
  {Gaberdiel}, \citenamefont {Gopakumar},\ and\ \citenamefont
  {Nairz}}]{Gaberdiel:2023lco}%
  \BibitemOpen
  \bibfield  {author} {\bibinfo {author} {\bibfnamefont {M.~R.}\ \bibnamefont
  {Gaberdiel}}, \bibinfo {author} {\bibfnamefont {R.}~\bibnamefont
  {Gopakumar}},\ and\ \bibinfo {author} {\bibfnamefont {B.}~\bibnamefont
  {Nairz}},\ }\href {https://doi.org/10.1007/JHEP06(2024)030} {\bibfield
  {journal} {\bibinfo  {journal} {JHEP}\ }\textbf {\bibinfo {volume} {06}},\
  \bibinfo {pages} {030}},\ \Eprint {https://arxiv.org/abs/2312.13288}
  {arXiv:2312.13288 [hep-th]} \BibitemShut {NoStop}%
\bibitem [{\citenamefont {Gaberdiel}\ \emph
  {et~al.}(2025{\natexlab{a}})\citenamefont {Gaberdiel}, \citenamefont
  {Lichtner},\ and\ \citenamefont {Nairz}}]{Gaberdiel:2024nge}%
  \BibitemOpen
  \bibfield  {author} {\bibinfo {author} {\bibfnamefont {M.~R.}\ \bibnamefont
  {Gaberdiel}}, \bibinfo {author} {\bibfnamefont {F.}~\bibnamefont
  {Lichtner}},\ and\ \bibinfo {author} {\bibfnamefont {B.}~\bibnamefont
  {Nairz}},\ }\href {https://doi.org/10.1007/JHEP05(2025)084} {\bibfield
  {journal} {\bibinfo  {journal} {JHEP}\ }\textbf {\bibinfo {volume} {05}},\
  \bibinfo {pages} {084}},\ \Eprint {https://arxiv.org/abs/2411.17612}
  {arXiv:2411.17612 [hep-th]} \BibitemShut {NoStop}%
\bibitem [{\citenamefont {Gaberdiel}\ \emph
  {et~al.}(2025{\natexlab{b}})\citenamefont {Gaberdiel}, \citenamefont
  {Kempel},\ and\ \citenamefont {Nairz}}]{Gaberdiel:2024dfw}%
  \BibitemOpen
  \bibfield  {author} {\bibinfo {author} {\bibfnamefont {M.~R.}\ \bibnamefont
  {Gaberdiel}}, \bibinfo {author} {\bibfnamefont {D.}~\bibnamefont {Kempel}},\
  and\ \bibinfo {author} {\bibfnamefont {B.}~\bibnamefont {Nairz}},\ }\href
  {https://doi.org/10.1007/JHEP05(2025)229} {\bibfield  {journal} {\bibinfo
  {journal} {JHEP}\ }\textbf {\bibinfo {volume} {05}},\ \bibinfo {pages}
  {229}},\ \Eprint {https://arxiv.org/abs/2412.02741} {arXiv:2412.02741
  [hep-th]} \BibitemShut {NoStop}%
\bibitem [{\citenamefont {Berenstein}\ \emph {et~al.}(2002)\citenamefont
  {Berenstein}, \citenamefont {Maldacena},\ and\ \citenamefont
  {Nastase}}]{Berenstein:2002jq}%
  \BibitemOpen
  \bibfield  {author} {\bibinfo {author} {\bibfnamefont {D.~E.}\ \bibnamefont
  {Berenstein}}, \bibinfo {author} {\bibfnamefont {J.~M.}\ \bibnamefont
  {Maldacena}},\ and\ \bibinfo {author} {\bibfnamefont {H.~S.}\ \bibnamefont
  {Nastase}},\ }\href {https://doi.org/10.1088/1126-6708/2002/04/013}
  {\bibfield  {journal} {\bibinfo  {journal} {JHEP}\ }\textbf {\bibinfo
  {volume} {04}},\ \bibinfo {pages} {013}},\ \Eprint
  {https://arxiv.org/abs/hep-th/0202021} {arXiv:hep-th/0202021} \BibitemShut
  {NoStop}%
\bibitem [{\citenamefont {Beisert}\ \emph {et~al.}(2003)\citenamefont
  {Beisert}, \citenamefont {Kristjansen},\ and\ \citenamefont
  {Staudacher}}]{Beisert:2003tq}%
  \BibitemOpen
  \bibfield  {author} {\bibinfo {author} {\bibfnamefont {N.}~\bibnamefont
  {Beisert}}, \bibinfo {author} {\bibfnamefont {C.}~\bibnamefont
  {Kristjansen}},\ and\ \bibinfo {author} {\bibfnamefont {M.}~\bibnamefont
  {Staudacher}},\ }\href {https://doi.org/10.1016/S0550-3213(03)00406-1}
  {\bibfield  {journal} {\bibinfo  {journal} {Nucl. Phys. B}\ }\textbf
  {\bibinfo {volume} {664}},\ \bibinfo {pages} {131} (\bibinfo {year}
  {2003})},\ \Eprint {https://arxiv.org/abs/hep-th/0303060}
  {arXiv:hep-th/0303060} \BibitemShut {NoStop}%
\bibitem [{Note1()}]{Note1}%
  \BibitemOpen
  \bibinfo {note} {The coupling constant $g_s$ of the dual AdS$_3$ string
  theory is proportional to $g_s\sim 1/\protect \sqrt {N}$. However, the
  subleading anomalous conformal dimensions involve the splitting and joining
  of strings, see FIG.~\ref {fig:cov_surf}, and hence are suppressed by $g_s^2
  \sim 1/N$.}\BibitemShut {Stop}%
\bibitem [{Note2()}]{Note2}%
  \BibitemOpen
  \bibinfo {note} {In a companion paper \cite {PRDcomp} we have studied the
  behavior of the degeneracies to subleading order in $\protect \frac {1}{N}$,
  and found direct evidence that integrability is indeed broken by these
  effects.}\BibitemShut {Stop}%
\bibitem [{\citenamefont {Dixon}(1987)}]{Dixon:1987bg}%
  \BibitemOpen
  \bibfield  {author} {\bibinfo {author} {\bibfnamefont {L.~J.}\ \bibnamefont
  {Dixon}},\ }in\ \href@noop {} {\emph {\bibinfo {booktitle} {{Summer Workshop
  in High-energy Physics and Cosmology}}}}\ (\bibinfo {year}
  {1987})\BibitemShut {NoStop}%
\bibitem [{\citenamefont {Pakman}\ \emph {et~al.}(2009)\citenamefont {Pakman},
  \citenamefont {Rastelli},\ and\ \citenamefont {Razamat}}]{Pakman:2009zz}%
  \BibitemOpen
  \bibfield  {author} {\bibinfo {author} {\bibfnamefont {A.}~\bibnamefont
  {Pakman}}, \bibinfo {author} {\bibfnamefont {L.}~\bibnamefont {Rastelli}},\
  and\ \bibinfo {author} {\bibfnamefont {S.~S.}\ \bibnamefont {Razamat}},\
  }\href {https://doi.org/10.1088/1126-6708/2009/10/034} {\bibfield  {journal}
  {\bibinfo  {journal} {JHEP}\ }\textbf {\bibinfo {volume} {10}},\ \bibinfo
  {pages} {034}},\ \Eprint {https://arxiv.org/abs/0905.3448} {arXiv:0905.3448
  [hep-th]} \BibitemShut {NoStop}%
\bibitem [{Note3()}]{Note3}%
  \BibitemOpen
  \bibinfo {note} {Note that the expansions in $g$ and $1/N$ have distinct
  origins: $g$ is the coupling constant of the perturbation by $\mathinner
  {|{\Phi }\rangle }$, and to lowest non-trivial order, the anomalous conformal
  dimensions appear at second order in the perturbation, i.e.\ are proportional
  to $g^2$. On the other hand, the factors of $N$ come from the combinatorics
  of the contractions and distinguish, in particular, single-cycle and
  double-cycle intermediate states.}\BibitemShut {Stop}%
\bibitem [{\citenamefont {Gaberdiel}\ \emph {et~al.}(2026)\citenamefont
  {Gaberdiel}, \citenamefont {Nairz},\ and\ \citenamefont {Peng}}]{PRDcomp}%
  \BibitemOpen
  \bibfield  {author} {\bibinfo {author} {\bibfnamefont {M.~R.}\ \bibnamefont
  {Gaberdiel}}, \bibinfo {author} {\bibfnamefont {B.}~\bibnamefont {Nairz}},\
  and\ \bibinfo {author} {\bibfnamefont {C.}~\bibnamefont {Peng}},\ }\href@noop
  {} {\  (\bibinfo {year} {2026})}\BibitemShut {NoStop}%
\bibitem [{\citenamefont {Lunin}\ and\ \citenamefont
  {Mathur}(2001)}]{Lunin:2000yv}%
  \BibitemOpen
  \bibfield  {author} {\bibinfo {author} {\bibfnamefont {O.}~\bibnamefont
  {Lunin}}\ and\ \bibinfo {author} {\bibfnamefont {S.~D.}\ \bibnamefont
  {Mathur}},\ }\href {https://doi.org/10.1007/s002200100431} {\bibfield
  {journal} {\bibinfo  {journal} {Commun. Math. Phys.}\ }\textbf {\bibinfo
  {volume} {219}},\ \bibinfo {pages} {399} (\bibinfo {year} {2001})},\ \Eprint
  {https://arxiv.org/abs/hep-th/0006196} {arXiv:hep-th/0006196} \BibitemShut
  {NoStop}%
\bibitem [{\citenamefont {Haake}(2010)}]{Haake:2010fgh}%
  \BibitemOpen
  \bibfield  {author} {\bibinfo {author} {\bibfnamefont {F.}~\bibnamefont
  {Haake}},\ }\href {https://doi.org/10.1007/978-3-642-05428-0} {\emph
  {\bibinfo {title} {{Quantum Signatures of Chaos}}}},\ Springer Series in
  Synergetics\ (\bibinfo  {publisher} {Springer},\ \bibinfo {address}
  {Berlin},\ \bibinfo {year} {2010})\BibitemShut {NoStop}%
\bibitem [{\citenamefont {Sekino}\ and\ \citenamefont
  {Susskind}(2008)}]{Sekino:2008he}%
  \BibitemOpen
  \bibfield  {author} {\bibinfo {author} {\bibfnamefont {Y.}~\bibnamefont
  {Sekino}}\ and\ \bibinfo {author} {\bibfnamefont {L.}~\bibnamefont
  {Susskind}},\ }\href {https://doi.org/10.1088/1126-6708/2008/10/065}
  {\bibfield  {journal} {\bibinfo  {journal} {JHEP}\ }\textbf {\bibinfo
  {volume} {10}},\ \bibinfo {pages} {065}},\ \Eprint
  {https://arxiv.org/abs/0808.2096} {arXiv:0808.2096 [hep-th]} \BibitemShut
  {NoStop}%
\bibitem [{\citenamefont {Shenker}\ and\ \citenamefont
  {Stanford}(2014)}]{Shenker:2013pqa}%
  \BibitemOpen
  \bibfield  {author} {\bibinfo {author} {\bibfnamefont {S.~H.}\ \bibnamefont
  {Shenker}}\ and\ \bibinfo {author} {\bibfnamefont {D.}~\bibnamefont
  {Stanford}},\ }\href {https://doi.org/10.1007/JHEP03(2014)067} {\bibfield
  {journal} {\bibinfo  {journal} {JHEP}\ }\textbf {\bibinfo {volume} {03}},\
  \bibinfo {pages} {067}},\ \Eprint {https://arxiv.org/abs/1306.0622}
  {arXiv:1306.0622 [hep-th]} \BibitemShut {NoStop}%
\bibitem [{\citenamefont {Kitaev}(2015)}]{kitaev2015simple}%
  \BibitemOpen
  \bibfield  {author} {\bibinfo {author} {\bibfnamefont {A.}~\bibnamefont
  {Kitaev}},\ }in\ \href@noop {} {\emph {\bibinfo {booktitle} {KITP strings
  seminar and Entanglement}}},\ Vol.~\bibinfo {volume} {12}\ (\bibinfo {year}
  {2015})\BibitemShut {NoStop}%
\bibitem [{\citenamefont {Maldacena}\ and\ \citenamefont
  {Stanford}(2016)}]{Maldacena:2016hyu}%
  \BibitemOpen
  \bibfield  {author} {\bibinfo {author} {\bibfnamefont {J.}~\bibnamefont
  {Maldacena}}\ and\ \bibinfo {author} {\bibfnamefont {D.}~\bibnamefont
  {Stanford}},\ }\href {https://doi.org/10.1103/PhysRevD.94.106002} {\bibfield
  {journal} {\bibinfo  {journal} {Phys. Rev. D}\ }\textbf {\bibinfo {volume}
  {94}},\ \bibinfo {pages} {106002} (\bibinfo {year} {2016})},\ \Eprint
  {https://arxiv.org/abs/1604.07818} {arXiv:1604.07818 [hep-th]} \BibitemShut
  {NoStop}%
\bibitem [{\citenamefont {You}\ \emph {et~al.}(2017)\citenamefont {You},
  \citenamefont {Ludwig},\ and\ \citenamefont {Xu}}]{You:2016ldz}%
  \BibitemOpen
  \bibfield  {author} {\bibinfo {author} {\bibfnamefont {Y.-Z.}\ \bibnamefont
  {You}}, \bibinfo {author} {\bibfnamefont {A.~W.~W.}\ \bibnamefont {Ludwig}},\
  and\ \bibinfo {author} {\bibfnamefont {C.}~\bibnamefont {Xu}},\ }\href
  {https://doi.org/10.1103/PhysRevB.95.115150} {\bibfield  {journal} {\bibinfo
  {journal} {Phys. Rev. B}\ }\textbf {\bibinfo {volume} {95}},\ \bibinfo
  {pages} {115150} (\bibinfo {year} {2017})},\ \Eprint
  {https://arxiv.org/abs/1602.06964} {arXiv:1602.06964 [cond-mat.str-el]}
  \BibitemShut {NoStop}%
\bibitem [{\citenamefont {Garc{\'\i}a-Garc{\'\i}a}\ and\ \citenamefont
  {Verbaarschot}(2016)}]{Garcia-Garcia:2016mno}%
  \BibitemOpen
  \bibfield  {author} {\bibinfo {author} {\bibfnamefont {A.~M.}\ \bibnamefont
  {Garc{\'\i}a-Garc{\'\i}a}}\ and\ \bibinfo {author} {\bibfnamefont {J.~J.~M.}\
  \bibnamefont {Verbaarschot}},\ }\href
  {https://doi.org/10.1103/PhysRevD.94.126010} {\bibfield  {journal} {\bibinfo
  {journal} {Phys. Rev. D}\ }\textbf {\bibinfo {volume} {94}},\ \bibinfo
  {pages} {126010} (\bibinfo {year} {2016})},\ \Eprint
  {https://arxiv.org/abs/1610.03816} {arXiv:1610.03816 [hep-th]} \BibitemShut
  {NoStop}%
\bibitem [{\citenamefont {Cotler}\ \emph {et~al.}(2017)\citenamefont {Cotler},
  \citenamefont {Gur-Ari}, \citenamefont {Hanada}, \citenamefont {Polchinski},
  \citenamefont {Saad}, \citenamefont {Shenker}, \citenamefont {Stanford},
  \citenamefont {Streicher},\ and\ \citenamefont {Tezuka}}]{Cotler:2016fpe}%
  \BibitemOpen
  \bibfield  {author} {\bibinfo {author} {\bibfnamefont {J.~S.}\ \bibnamefont
  {Cotler}}, \bibinfo {author} {\bibfnamefont {G.}~\bibnamefont {Gur-Ari}},
  \bibinfo {author} {\bibfnamefont {M.}~\bibnamefont {Hanada}}, \bibinfo
  {author} {\bibfnamefont {J.}~\bibnamefont {Polchinski}}, \bibinfo {author}
  {\bibfnamefont {P.}~\bibnamefont {Saad}}, \bibinfo {author} {\bibfnamefont
  {S.~H.}\ \bibnamefont {Shenker}}, \bibinfo {author} {\bibfnamefont
  {D.}~\bibnamefont {Stanford}}, \bibinfo {author} {\bibfnamefont
  {A.}~\bibnamefont {Streicher}},\ and\ \bibinfo {author} {\bibfnamefont
  {M.}~\bibnamefont {Tezuka}},\ }\href
  {https://doi.org/10.1007/JHEP05(2017)118} {\bibfield  {journal} {\bibinfo
  {journal} {JHEP}\ }\textbf {\bibinfo {volume} {05}},\ \bibinfo {pages}
  {118}},\ \bibinfo {note} {[Erratum: JHEP 09, 002 (2018)]},\ \Eprint
  {https://arxiv.org/abs/1611.04650} {arXiv:1611.04650 [hep-th]} \BibitemShut
  {NoStop}%
\bibitem [{\citenamefont {Stanford}\ and\ \citenamefont
  {Witten}(2020)}]{Stanford:2019vob}%
  \BibitemOpen
  \bibfield  {author} {\bibinfo {author} {\bibfnamefont {D.}~\bibnamefont
  {Stanford}}\ and\ \bibinfo {author} {\bibfnamefont {E.}~\bibnamefont
  {Witten}},\ }\href {https://doi.org/10.4310/ATMP.2020.v24.n6.a4} {\bibfield
  {journal} {\bibinfo  {journal} {Adv. Theor. Math. Phys.}\ }\textbf {\bibinfo
  {volume} {24}},\ \bibinfo {pages} {1475} (\bibinfo {year} {2020})},\ \Eprint
  {https://arxiv.org/abs/1907.03363} {arXiv:1907.03363 [hep-th]} \BibitemShut
  {NoStop}%
\bibitem [{\citenamefont {Witten}(2020)}]{Witten:2020wvy}%
  \BibitemOpen
  \bibfield  {author} {\bibinfo {author} {\bibfnamefont {E.}~\bibnamefont
  {Witten}},\ }\href {https://doi.org/10.1098/rspa.2020.0582} {\bibfield
  {journal} {\bibinfo  {journal} {Proc. Roy. Soc. Lond. A}\ }\textbf {\bibinfo
  {volume} {476}},\ \bibinfo {pages} {20200582} (\bibinfo {year} {2020})},\
  \Eprint {https://arxiv.org/abs/2006.13414} {arXiv:2006.13414 [hep-th]}
  \BibitemShut {NoStop}%
\bibitem [{\citenamefont {Turiaci}\ and\ \citenamefont
  {Witten}(2023)}]{Turiaci:2023jfa}%
  \BibitemOpen
  \bibfield  {author} {\bibinfo {author} {\bibfnamefont {G.~J.}\ \bibnamefont
  {Turiaci}}\ and\ \bibinfo {author} {\bibfnamefont {E.}~\bibnamefont
  {Witten}},\ }\href {https://doi.org/10.1007/JHEP12(2023)003} {\bibfield
  {journal} {\bibinfo  {journal} {JHEP}\ }\textbf {\bibinfo {volume} {12}},\
  \bibinfo {pages} {003}},\ \Eprint {https://arxiv.org/abs/2305.19438}
  {arXiv:2305.19438 [hep-th]} \BibitemShut {NoStop}%
\bibitem [{\citenamefont {McLoughlin}\ \emph {et~al.}(2023)\citenamefont
  {McLoughlin}, \citenamefont {Pereira},\ and\ \citenamefont
  {Spiering}}]{McLoughlin:2020zew}%
  \BibitemOpen
  \bibfield  {author} {\bibinfo {author} {\bibfnamefont {T.}~\bibnamefont
  {McLoughlin}}, \bibinfo {author} {\bibfnamefont {R.}~\bibnamefont
  {Pereira}},\ and\ \bibinfo {author} {\bibfnamefont {A.}~\bibnamefont
  {Spiering}},\ }\href {https://doi.org/10.21468/SciPostPhys.14.3.049}
  {\bibfield  {journal} {\bibinfo  {journal} {SciPost Phys.}\ }\textbf
  {\bibinfo {volume} {14}},\ \bibinfo {pages} {049} (\bibinfo {year} {2023})},\
  \Eprint {https://arxiv.org/abs/2011.04633} {arXiv:2011.04633 [hep-th]}
  \BibitemShut {NoStop}%
\bibitem [{\citenamefont {Kristjansen}\ \emph {et~al.}(2009)\citenamefont
  {Kristjansen}, \citenamefont {Orselli},\ and\ \citenamefont
  {Zoubos}}]{Kristjansen:2008ABJM}%
  \BibitemOpen
  \bibfield  {author} {\bibinfo {author} {\bibfnamefont {C.}~\bibnamefont
  {Kristjansen}}, \bibinfo {author} {\bibfnamefont {M.}~\bibnamefont
  {Orselli}},\ and\ \bibinfo {author} {\bibfnamefont {K.}~\bibnamefont
  {Zoubos}},\ }\href {https://doi.org/10.1088/1126-6708/2009/03/037} {\bibfield
   {journal} {\bibinfo  {journal} {JHEP}\ }\textbf {\bibinfo {volume} {03}},\
  \bibinfo {pages} {037}},\ \Eprint {https://arxiv.org/abs/0811.2150}
  {arXiv:0811.2150 [hep-th]} \BibitemShut {NoStop}%
\bibitem [{\citenamefont {Guhr}\ \emph {et~al.}(1998)\citenamefont {Guhr},
  \citenamefont {Muller-Groeling},\ and\ \citenamefont
  {Weidenmuller}}]{Guhr:1997ve}%
  \BibitemOpen
  \bibfield  {author} {\bibinfo {author} {\bibfnamefont {T.}~\bibnamefont
  {Guhr}}, \bibinfo {author} {\bibfnamefont {A.}~\bibnamefont
  {Muller-Groeling}},\ and\ \bibinfo {author} {\bibfnamefont {H.~A.}\
  \bibnamefont {Weidenmuller}},\ }\href
  {https://doi.org/10.1016/S0370-1573(97)00088-4} {\bibfield  {journal}
  {\bibinfo  {journal} {Phys. Rept.}\ }\textbf {\bibinfo {volume} {299}},\
  \bibinfo {pages} {189} (\bibinfo {year} {1998})},\ \Eprint
  {https://arxiv.org/abs/cond-mat/9707301} {arXiv:cond-mat/9707301}
  \BibitemShut {NoStop}%
\bibitem [{Note4()}]{Note4}%
  \BibitemOpen
  \bibinfo {note} {For spectra with slowly varying average density, the gap
  ratio is expected to be independent of unfolding. We have checked that
  $\protect \langle r \protect \rangle $ varies by less than $10^{-4}$ when
  using raw or unfolded eigenvalues, suggesting that the unfolding procedure is
  consistent.}\BibitemShut {Stop}%
\bibitem [{\citenamefont {Atas}\ \emph {et~al.}(2013)\citenamefont {Atas},
  \citenamefont {Bogomolny}, \citenamefont {Giraud},\ and\ \citenamefont
  {Roux}}]{Atas:2013gvn}%
  \BibitemOpen
  \bibfield  {author} {\bibinfo {author} {\bibfnamefont {Y.~Y.}\ \bibnamefont
  {Atas}}, \bibinfo {author} {\bibfnamefont {E.}~\bibnamefont {Bogomolny}},
  \bibinfo {author} {\bibfnamefont {O.}~\bibnamefont {Giraud}},\ and\ \bibinfo
  {author} {\bibfnamefont {G.}~\bibnamefont {Roux}},\ }\href
  {https://doi.org/10.1103/PhysRevLett.110.084101} {\bibfield  {journal}
  {\bibinfo  {journal} {Phys. Rev. Lett.}\ }\textbf {\bibinfo {volume} {110}},\
  \bibinfo {pages} {084101} (\bibinfo {year} {2013})}\BibitemShut {NoStop}%
\bibitem [{Note5()}]{Note5}%
  \BibitemOpen
  \bibinfo {note} {For the planar eigenvalues, we implement a block bootstrap,
  realised by partitioning the gap ratios into blocks from which one randomly
  chooses with replacement. For the non-planar data, we instead implement a
  cluster bootstrap, by randomly drawing from the smaller $\protect \tilde
  {\gamma }_1|_{\protect \mathcal {E}}$ matrices.}\BibitemShut {Stop}%
\bibitem [{Note6()}]{Note6}%
  \BibitemOpen
  \bibinfo {note} {There have been previous observations of transitions between
  chaotic and integrable behavior in many examples of holographic disordered
  models \cite {Peng:2018zap,Chang:2021fmd,Chang:2021wbx,Berkooz:2022dfr,
  Berkooz:2024evs, Gao:2024lve}. These obervations suggest that large
  symmetries or integrable structures can emerge at special points in the
  moduli space of chaotic theories. Interestingly, non-maximal chaotic behavior
  was found for certain 2D and 3D disordered models in~\cite
  {Peng:2018zap,Chang:2021wbx}.}\BibitemShut {Stop}%
\bibitem [{Note7()}]{Note7}%
  \BibitemOpen
  \bibinfo {note} {Notice that the appearance of chaotic behavior in some BPS
  sector could be related to recent studies of BPS chaos \cite {Chen:2024oqv},
  diagnosed by the LMRS prescription \cite {Lin:2022rzw,Lin:2022zxd} or
  recently also by a Berry curvature \cite {Chen:2026vml}, and the fortuity
  phenomena \cite {Chang:2013fba, Chang:2022mjp, Chang:2024lxt} especially in
  the orbifold CFT setup~\cite {Chang:2025rqy,Chang:2025wgo}.}\BibitemShut
  {Stop}%
\bibitem [{\citenamefont {Beisert}\ \emph {et~al.}(2012)\citenamefont {Beisert}
  \emph {et~al.}}]{Beisert:2010jr}%
  \BibitemOpen
  \bibfield  {author} {\bibinfo {author} {\bibfnamefont {N.}~\bibnamefont
  {Beisert}} \emph {et~al.},\ }\href
  {https://doi.org/10.1007/s11005-011-0529-2} {\bibfield  {journal} {\bibinfo
  {journal} {Lett. Math. Phys.}\ }\textbf {\bibinfo {volume} {99}},\ \bibinfo
  {pages} {3} (\bibinfo {year} {2012})},\ \Eprint
  {https://arxiv.org/abs/1012.3982} {arXiv:1012.3982 [hep-th]} \BibitemShut
  {NoStop}%
\bibitem [{Note8()}]{Note8}%
  \BibitemOpen
  \bibinfo {note} {In particular, the symmetric orbifold theory seems to be
  vector-like (rather than a gauge-theory) in that the central charge is
  proportional to $N$, rather than $N^2$.}\BibitemShut {Stop}%
\bibitem [{\citenamefont {Maldacena}\ and\ \citenamefont
  {Strominger}(1998)}]{Maldacena:1998bw}%
  \BibitemOpen
  \bibfield  {author} {\bibinfo {author} {\bibfnamefont {J.~M.}\ \bibnamefont
  {Maldacena}}\ and\ \bibinfo {author} {\bibfnamefont {A.}~\bibnamefont
  {Strominger}},\ }\href {https://doi.org/10.1088/1126-6708/1998/12/005}
  {\bibfield  {journal} {\bibinfo  {journal} {JHEP}\ }\textbf {\bibinfo
  {volume} {12}},\ \bibinfo {pages} {005}},\ \Eprint
  {https://arxiv.org/abs/hep-th/9804085} {arXiv:hep-th/9804085} \BibitemShut
  {NoStop}%
\bibitem [{Note9()}]{Note9}%
  \BibitemOpen
  \bibinfo {note} {An illustrative example of this paradigm is the
  Sachdev-Ye-Kitaev model~\cite {Sachdev:1992fk, kitaev2015simple}, where the
  light states near the spectral edge carry additional structure of the
  Schwarzian sector that dominates at low temperatures and maps directly to
  Jackiw-Teitelboim gravity~\cite {Maldacena:2016hyu,Maldacena:2016upp,
  Garcia-Garcia:2016mno}.}\BibitemShut {Stop}%
\bibitem [{\citenamefont {Zhang}(2026)}]{Zhang:2026jnf}%
  \BibitemOpen
  \bibfield  {author} {\bibinfo {author} {\bibfnamefont {H.}~\bibnamefont
  {Zhang}},\ }\href@noop {} {\  (\bibinfo {year} {2026})},\ \Eprint
  {https://arxiv.org/abs/2605.18725} {arXiv:2605.18725 [hep-th]} \BibitemShut
  {NoStop}%
\bibitem [{\citenamefont {Peng}(2018)}]{Peng:2018zap}%
  \BibitemOpen
  \bibfield  {author} {\bibinfo {author} {\bibfnamefont {C.}~\bibnamefont
  {Peng}},\ }\href {https://doi.org/10.1007/JHEP12(2018)065} {\bibfield
  {journal} {\bibinfo  {journal} {JHEP}\ }\textbf {\bibinfo {volume} {12}},\
  \bibinfo {pages} {065}},\ \Eprint {https://arxiv.org/abs/1805.09325}
  {arXiv:1805.09325 [hep-th]} \BibitemShut {NoStop}%
\bibitem [{\citenamefont {Chang}\ \emph {et~al.}(2021)\citenamefont {Chang},
  \citenamefont {Colin-Ellerin}, \citenamefont {Peng},\ and\ \citenamefont
  {Rangamani}}]{Chang:2021fmd}%
  \BibitemOpen
  \bibfield  {author} {\bibinfo {author} {\bibfnamefont {C.-M.}\ \bibnamefont
  {Chang}}, \bibinfo {author} {\bibfnamefont {S.}~\bibnamefont
  {Colin-Ellerin}}, \bibinfo {author} {\bibfnamefont {C.}~\bibnamefont
  {Peng}},\ and\ \bibinfo {author} {\bibfnamefont {M.}~\bibnamefont
  {Rangamani}},\ }\href {https://doi.org/10.1007/JHEP11(2021)211} {\bibfield
  {journal} {\bibinfo  {journal} {JHEP}\ }\textbf {\bibinfo {volume} {11}},\
  \bibinfo {pages} {211}},\ \Eprint {https://arxiv.org/abs/2108.00027}
  {arXiv:2108.00027 [hep-th]} \BibitemShut {NoStop}%
\bibitem [{\citenamefont {Chang}\ \emph {et~al.}(2022)\citenamefont {Chang},
  \citenamefont {Colin-Ellerin}, \citenamefont {Peng},\ and\ \citenamefont
  {Rangamani}}]{Chang:2021wbx}%
  \BibitemOpen
  \bibfield  {author} {\bibinfo {author} {\bibfnamefont {C.-M.}\ \bibnamefont
  {Chang}}, \bibinfo {author} {\bibfnamefont {S.}~\bibnamefont
  {Colin-Ellerin}}, \bibinfo {author} {\bibfnamefont {C.}~\bibnamefont
  {Peng}},\ and\ \bibinfo {author} {\bibfnamefont {M.}~\bibnamefont
  {Rangamani}},\ }\href {https://doi.org/10.1103/PhysRevLett.129.011603}
  {\bibfield  {journal} {\bibinfo  {journal} {Phys. Rev. Lett.}\ }\textbf
  {\bibinfo {volume} {129}},\ \bibinfo {pages} {011603} (\bibinfo {year}
  {2022})},\ \Eprint {https://arxiv.org/abs/2112.09157} {arXiv:2112.09157
  [hep-th]} \BibitemShut {NoStop}%
\bibitem [{\citenamefont {Berkooz}\ \emph {et~al.}(2022)\citenamefont
  {Berkooz}, \citenamefont {Sharon}, \citenamefont {Silberstein},\ and\
  \citenamefont {Urbach}}]{Berkooz:2022dfr}%
  \BibitemOpen
  \bibfield  {author} {\bibinfo {author} {\bibfnamefont {M.}~\bibnamefont
  {Berkooz}}, \bibinfo {author} {\bibfnamefont {A.}~\bibnamefont {Sharon}},
  \bibinfo {author} {\bibfnamefont {N.}~\bibnamefont {Silberstein}},\ and\
  \bibinfo {author} {\bibfnamefont {E.~Y.}\ \bibnamefont {Urbach}},\ }\href
  {https://doi.org/10.1103/PhysRevLett.129.071601} {\bibfield  {journal}
  {\bibinfo  {journal} {Phys. Rev. Lett.}\ }\textbf {\bibinfo {volume} {129}},\
  \bibinfo {pages} {071601} (\bibinfo {year} {2022})},\ \Eprint
  {https://arxiv.org/abs/2207.11980} {arXiv:2207.11980 [hep-th]} \BibitemShut
  {NoStop}%
\bibitem [{\citenamefont {Berkooz}\ \emph {et~al.}(2024)\citenamefont
  {Berkooz}, \citenamefont {Brukner}, \citenamefont {Jia},\ and\ \citenamefont
  {Mamroud}}]{Berkooz:2024evs}%
  \BibitemOpen
  \bibfield  {author} {\bibinfo {author} {\bibfnamefont {M.}~\bibnamefont
  {Berkooz}}, \bibinfo {author} {\bibfnamefont {N.}~\bibnamefont {Brukner}},
  \bibinfo {author} {\bibfnamefont {Y.}~\bibnamefont {Jia}},\ and\ \bibinfo
  {author} {\bibfnamefont {O.}~\bibnamefont {Mamroud}},\ }\href
  {https://doi.org/10.1103/PhysRevLett.133.221602} {\bibfield  {journal}
  {\bibinfo  {journal} {Phys. Rev. Lett.}\ }\textbf {\bibinfo {volume} {133}},\
  \bibinfo {pages} {221602} (\bibinfo {year} {2024})},\ \Eprint
  {https://arxiv.org/abs/2403.01950} {arXiv:2403.01950 [hep-th]} \BibitemShut
  {NoStop}%
\bibitem [{\citenamefont {Gao}\ \emph {et~al.}(2025)\citenamefont {Gao},
  \citenamefont {Lin},\ and\ \citenamefont {Peng}}]{Gao:2024lve}%
  \BibitemOpen
  \bibfield  {author} {\bibinfo {author} {\bibfnamefont {P.}~\bibnamefont
  {Gao}}, \bibinfo {author} {\bibfnamefont {H.}~\bibnamefont {Lin}},\ and\
  \bibinfo {author} {\bibfnamefont {C.}~\bibnamefont {Peng}},\ }\href
  {https://doi.org/10.1103/1k6k-xhmk} {\bibfield  {journal} {\bibinfo
  {journal} {Phys. Rev. D}\ }\textbf {\bibinfo {volume} {112}},\ \bibinfo
  {pages} {066012} (\bibinfo {year} {2025})},\ \Eprint
  {https://arxiv.org/abs/2411.12806} {arXiv:2411.12806 [hep-th]} \BibitemShut
  {NoStop}%
\bibitem [{\citenamefont {Chen}\ \emph {et~al.}(2025)\citenamefont {Chen},
  \citenamefont {Lin},\ and\ \citenamefont {Shenker}}]{Chen:2024oqv}%
  \BibitemOpen
  \bibfield  {author} {\bibinfo {author} {\bibfnamefont {Y.}~\bibnamefont
  {Chen}}, \bibinfo {author} {\bibfnamefont {H.~W.}\ \bibnamefont {Lin}},\ and\
  \bibinfo {author} {\bibfnamefont {S.~H.}\ \bibnamefont {Shenker}},\ }\href
  {https://doi.org/10.21468/SciPostPhys.18.2.072} {\bibfield  {journal}
  {\bibinfo  {journal} {SciPost Phys.}\ }\textbf {\bibinfo {volume} {18}},\
  \bibinfo {pages} {072} (\bibinfo {year} {2025})},\ \Eprint
  {https://arxiv.org/abs/2407.19387} {arXiv:2407.19387 [hep-th]} \BibitemShut
  {NoStop}%
\bibitem [{\citenamefont {Lin}\ \emph {et~al.}(2023{\natexlab{a}})\citenamefont
  {Lin}, \citenamefont {Maldacena}, \citenamefont {Rozenberg},\ and\
  \citenamefont {Shan}}]{Lin:2022rzw}%
  \BibitemOpen
  \bibfield  {author} {\bibinfo {author} {\bibfnamefont {H.~W.}\ \bibnamefont
  {Lin}}, \bibinfo {author} {\bibfnamefont {J.}~\bibnamefont {Maldacena}},
  \bibinfo {author} {\bibfnamefont {L.}~\bibnamefont {Rozenberg}},\ and\
  \bibinfo {author} {\bibfnamefont {J.}~\bibnamefont {Shan}},\ }\href
  {https://doi.org/10.21468/SciPostPhys.14.6.150} {\bibfield  {journal}
  {\bibinfo  {journal} {SciPost Phys.}\ }\textbf {\bibinfo {volume} {14}},\
  \bibinfo {pages} {150} (\bibinfo {year} {2023}{\natexlab{a}})},\ \Eprint
  {https://arxiv.org/abs/2207.00407} {arXiv:2207.00407 [hep-th]} \BibitemShut
  {NoStop}%
\bibitem [{\citenamefont {Lin}\ \emph {et~al.}(2023{\natexlab{b}})\citenamefont
  {Lin}, \citenamefont {Maldacena}, \citenamefont {Rozenberg},\ and\
  \citenamefont {Shan}}]{Lin:2022zxd}%
  \BibitemOpen
  \bibfield  {author} {\bibinfo {author} {\bibfnamefont {H.~W.}\ \bibnamefont
  {Lin}}, \bibinfo {author} {\bibfnamefont {J.}~\bibnamefont {Maldacena}},
  \bibinfo {author} {\bibfnamefont {L.}~\bibnamefont {Rozenberg}},\ and\
  \bibinfo {author} {\bibfnamefont {J.}~\bibnamefont {Shan}},\ }\href
  {https://doi.org/10.21468/SciPostPhys.14.5.128} {\bibfield  {journal}
  {\bibinfo  {journal} {SciPost Phys.}\ }\textbf {\bibinfo {volume} {14}},\
  \bibinfo {pages} {128} (\bibinfo {year} {2023}{\natexlab{b}})},\ \Eprint
  {https://arxiv.org/abs/2207.00408} {arXiv:2207.00408 [hep-th]} \BibitemShut
  {NoStop}%
\bibitem [{\citenamefont {Chen}\ \emph {et~al.}(2026)\citenamefont {Chen},
  \citenamefont {Colin-Ellerin}, \citenamefont {Mamroud},\ and\ \citenamefont
  {Papadodimas}}]{Chen:2026vml}%
  \BibitemOpen
  \bibfield  {author} {\bibinfo {author} {\bibfnamefont {Y.}~\bibnamefont
  {Chen}}, \bibinfo {author} {\bibfnamefont {S.}~\bibnamefont {Colin-Ellerin}},
  \bibinfo {author} {\bibfnamefont {O.}~\bibnamefont {Mamroud}},\ and\ \bibinfo
  {author} {\bibfnamefont {K.}~\bibnamefont {Papadodimas}},\ }\href@noop {} {\
  (\bibinfo {year} {2026})},\ \Eprint {https://arxiv.org/abs/2604.23287}
  {arXiv:2604.23287 [hep-th]} \BibitemShut {NoStop}%
\bibitem [{\citenamefont {Chang}\ and\ \citenamefont
  {Yin}(2013)}]{Chang:2013fba}%
  \BibitemOpen
  \bibfield  {author} {\bibinfo {author} {\bibfnamefont {C.-M.}\ \bibnamefont
  {Chang}}\ and\ \bibinfo {author} {\bibfnamefont {X.}~\bibnamefont {Yin}},\
  }\href {https://doi.org/10.1103/PhysRevD.88.106005} {\bibfield  {journal}
  {\bibinfo  {journal} {Phys. Rev. D}\ }\textbf {\bibinfo {volume} {88}},\
  \bibinfo {pages} {106005} (\bibinfo {year} {2013})},\ \Eprint
  {https://arxiv.org/abs/1305.6314} {arXiv:1305.6314 [hep-th]} \BibitemShut
  {NoStop}%
\bibitem [{\citenamefont {Chang}\ and\ \citenamefont
  {Lin}(2023)}]{Chang:2022mjp}%
  \BibitemOpen
  \bibfield  {author} {\bibinfo {author} {\bibfnamefont {C.-M.}\ \bibnamefont
  {Chang}}\ and\ \bibinfo {author} {\bibfnamefont {Y.-H.}\ \bibnamefont
  {Lin}},\ }\href {https://doi.org/10.1007/JHEP02(2023)109} {\bibfield
  {journal} {\bibinfo  {journal} {JHEP}\ }\textbf {\bibinfo {volume} {02}},\
  \bibinfo {pages} {109}},\ \Eprint {https://arxiv.org/abs/2209.06728}
  {arXiv:2209.06728 [hep-th]} \BibitemShut {NoStop}%
\bibitem [{\citenamefont {Chang}\ \emph
  {et~al.}(2025{\natexlab{a}})\citenamefont {Chang}, \citenamefont {Chen},
  \citenamefont {Sia},\ and\ \citenamefont {Yang}}]{Chang:2024lxt}%
  \BibitemOpen
  \bibfield  {author} {\bibinfo {author} {\bibfnamefont {C.-M.}\ \bibnamefont
  {Chang}}, \bibinfo {author} {\bibfnamefont {Y.}~\bibnamefont {Chen}},
  \bibinfo {author} {\bibfnamefont {B.~S.}\ \bibnamefont {Sia}},\ and\ \bibinfo
  {author} {\bibfnamefont {Z.}~\bibnamefont {Yang}},\ }\href
  {https://doi.org/10.1007/JHEP08(2025)003} {\bibfield  {journal} {\bibinfo
  {journal} {JHEP}\ }\textbf {\bibinfo {volume} {08}},\ \bibinfo {pages}
  {003}},\ \Eprint {https://arxiv.org/abs/2412.06902} {arXiv:2412.06902
  [hep-th]} \BibitemShut {NoStop}%
\bibitem [{\citenamefont {Chang}\ \emph
  {et~al.}(2025{\natexlab{b}})\citenamefont {Chang}, \citenamefont {Lin},\ and\
  \citenamefont {Zhang}}]{Chang:2025rqy}%
  \BibitemOpen
  \bibfield  {author} {\bibinfo {author} {\bibfnamefont {C.-M.}\ \bibnamefont
  {Chang}}, \bibinfo {author} {\bibfnamefont {Y.-H.}\ \bibnamefont {Lin}},\
  and\ \bibinfo {author} {\bibfnamefont {H.}~\bibnamefont {Zhang}},\
  }\href@noop {} {\  (\bibinfo {year} {2025}{\natexlab{b}})},\ \Eprint
  {https://arxiv.org/abs/2501.05448} {arXiv:2501.05448 [hep-th]} \BibitemShut
  {NoStop}%
\bibitem [{\citenamefont {Chang}\ and\ \citenamefont
  {Zhang}(2025)}]{Chang:2025wgo}%
  \BibitemOpen
  \bibfield  {author} {\bibinfo {author} {\bibfnamefont {C.-M.}\ \bibnamefont
  {Chang}}\ and\ \bibinfo {author} {\bibfnamefont {H.}~\bibnamefont {Zhang}},\
  }\href@noop {} {\  (\bibinfo {year} {2025})},\ \Eprint
  {https://arxiv.org/abs/2511.23294} {arXiv:2511.23294 [hep-th]} \BibitemShut
  {NoStop}%
\bibitem [{\citenamefont {Sachdev}\ and\ \citenamefont
  {Ye}(1993)}]{Sachdev:1992fk}%
  \BibitemOpen
  \bibfield  {author} {\bibinfo {author} {\bibfnamefont {S.}~\bibnamefont
  {Sachdev}}\ and\ \bibinfo {author} {\bibfnamefont {J.}~\bibnamefont {Ye}},\
  }\href {https://doi.org/10.1103/PhysRevLett.70.3339} {\bibfield  {journal}
  {\bibinfo  {journal} {Phys. Rev. Lett.}\ }\textbf {\bibinfo {volume} {70}},\
  \bibinfo {pages} {3339} (\bibinfo {year} {1993})},\ \Eprint
  {https://arxiv.org/abs/cond-mat/9212030} {arXiv:cond-mat/9212030}
  \BibitemShut {NoStop}%
\bibitem [{\citenamefont {Maldacena}\ \emph {et~al.}(2016)\citenamefont
  {Maldacena}, \citenamefont {Stanford},\ and\ \citenamefont
  {Yang}}]{Maldacena:2016upp}%
  \BibitemOpen
  \bibfield  {author} {\bibinfo {author} {\bibfnamefont {J.}~\bibnamefont
  {Maldacena}}, \bibinfo {author} {\bibfnamefont {D.}~\bibnamefont
  {Stanford}},\ and\ \bibinfo {author} {\bibfnamefont {Z.}~\bibnamefont
  {Yang}},\ }\href {https://doi.org/10.1093/ptep/ptw124} {\bibfield  {journal}
  {\bibinfo  {journal} {PTEP}\ }\textbf {\bibinfo {volume} {2016}},\ \bibinfo
  {pages} {12C104} (\bibinfo {year} {2016})},\ \Eprint
  {https://arxiv.org/abs/1606.01857} {arXiv:1606.01857 [hep-th]} \BibitemShut
  {NoStop}%
\end{thebibliography}%

\iftrue

\onecolumngrid
\pagebreak
\twocolumngrid
\begin{center}
\large\textbf{Appendix: Orbifold Calculation}
\end{center}

In this appendix we provide a technical discussion of the calculation of $\tilde{\gamma}_1$. We outline the determination of possible intermediate states and the covering map method for the calculation of string splitting processes.

Using Eq.~(\ref{eq:gamma_decomposition}), the non-planar correction $\tilde{\gamma}_1$ to the anomalous dimension matrix can be written as
\begin{align}\label{eq:gamma_decomposition_1}
\bra{\phi'}\tilde{\gamma}_1\ket{\phi} = \sum_{\substack{\chi\\ \text{multi-cycle}}} &\bra{\phi'}\tilde{G}'^+_{-\frac{1}{2}}\ket{\chi}_{\Phi} \bra{\chi}\tilde{G}'^-_{\frac{1}{2}}\ket{\phi}_{\Phi}  \\
&\quad + \bra{\phi'}\tilde{G}'^-_{\frac{1}{2}}\ket{\chi}_{\Phi}\bra{\chi}\tilde{G}'^+_{-\frac{1}{2}}\ket{\phi}_{\Phi}\ . \nonumber
\end{align}
Here, the sum ranges over states $\chi$ which lie in a multi-cycle twisted sector of cycle type $(w-k,k)$. The external states $\ket{\phi},\ket{\phi'}$, on the other hand, lie in a single-cycle twisted sector of length $w$. The quantities $\bra{\phi}\tilde{G}'^{\pm}_{\mp\frac{1}{2}}\ket{\chi}_{\Phi}$ denote regulated integrated three-point functions, defined in Eq.~(\ref{eq:regulated_transition}).
\medskip

\noindent The calculation of $\tilde{\gamma}_1$ thus broadly consists of two steps:
\begin{enumerate}
\item Find all intermediate multi-cycle states $\ket{\chi}$ which contribute to the sum in Eq.~(\ref{eq:gamma_decomposition_1}).
\item Evaluate the transition elements $\bra{\phi}\tilde{G}'^{\pm}_{\mp\frac{1}{2}}\ket{\chi}_{\Phi}$ for all external states $\ket{\phi},\ket{\phi'}$ and intermediate states $\ket{\chi}$.
\end{enumerate}
In the following we shall briefly discuss the key features of both steps.

\section{Intermediate states}

In this section, we discuss how to determine the intermediate states $\chi$ in Eq.~(\ref{eq:gamma_decomposition_1}) using the conserved charges of the deformation.

\subsection{Conserved charges}
The perturbation $\Phi$ in Eq.~(\ref{pert}) preserves the $\mcl{N}=(4,4)$ algebra of the orbifold, as well as several global symmetries associated to the $\mfr{so}(4)$ symmetry of the $\mbb{T}^4$ torus. In particular, the $\mfr{su}(2)$ charge measured by $K^3_0$ and two additional $\mfr{u}(1)$ charges $U_\text{bar}$, $U_\text{res}$ are conserved, see also \cite{Gaberdiel:2024nge}. Below, we list the dimensions and charges of the reference BPS states, as well as the two unprotected spaces $\mbb{V}_\mrm{b}$, $\mbb{V}_\mrm{f}$ of Eq.~(\ref{eq:reference_states}).
\begin{center}
\begin{tabular}{c|c|c|c|c}
 & $L_0$ & $K^3_0$ & $U_\text{bar}$ & $U_\text{res}$\\ \hline &&&\\[-8pt]
$\ket{w}$ & $\frac{w-1}{2}$ & $\frac{w-1}{2}$ & $0$ & $0$ \\[5pt]
$\ket{w-k}\otimes\ket{k}$ & $\frac{w-2}{2}$ & $\frac{w-2}{2}$ & $0$ & $0$ \\[5pt]
$\mbb{V}_\mrm{b}$ & $\frac{w+1}{2}$ & $\frac{w-1}{2}$ & $2$ & $2$ \\[5pt]
$\mbb{V}_\mrm{f}$ & $\frac{w-1}{2}$ & $\frac{w-3}{2}$ & $2$ & $0$
\end{tabular}
\end{center}

\subsection{Allowed transitions}

The transitions $\bra{\phi}\tilde{G}'^\pm_{\mp\frac{1}{2}}\ket{\chi}_\Phi$ are then constrained by charge conservation. The right-moving supercharges are uncharged w.r.t.~$L_0$, $K^3_0$, $U_\text{bar}$, but carry a $U_\text{res}$ charge of $\pm 1$. The matrix $\tilde{\gamma}_1$ then includes the transitions to all intermediate multi-cycle states $\chi$ with the correct charges. The number of such intermediate states is large.

For example, if we consider the state
\begin{equation}
\ket{\phi} := \alpha^1_{-\frac{m}{w}}\alpha^1_{-1+\frac{m}{w}}\ket{w}\ ,
\end{equation}
the allowed transitions under $\tilde{G}'^-_{\frac{1}{2}}$ include multi-cycle states
\begin{align}\label{eq:example_states}
\ket{\chi} &= \alpha^1_{-q}\psi^+_{-\frac{3}{2}+q}\ket{w-k}\otimes\ket{k}\ ,\nonumber\\ \ket{\chi}&=\alpha^1_{-q_1}\alpha^1_{-q_2}\bar\alpha^2_{-q_3}{\psi}^+_{-\frac{3}{2}+q_1+q_2+q_3}\ket{w-k}\otimes\ket{k}\ ,\nonumber\\
&\dots\ .
\end{align}
As $w$ increases, the number of these allowed intermediate states grows exponentially.

The fractional modes above can either act on the $w-k$ or $k$ cycle, which is reflected in the mode number $q_i$: in the former case, $q_i\in\frac{1}{w-k}\mbb{Z}$, while in the latter $q_i\in\frac{1}{k}\mbb{Z}$. In principle, different modes can act on different cycles---this needs to be taken into account when enumerating all possible states. We have automated the process of finding all intermediate states $\ket{\chi}$ with \texttt{Mathematica}.

\section{Covering map method}

Here, we describe the calculation of the quantities $\bra{\phi}\tilde{G}'^{\pm}_{\mp\frac{1}{2}}\ket{\chi}_{\Phi}$ entering Eq.~(\ref{eq:gamma_decomposition_1}).

\subsection{Transition amplitudes}

The transitions $\bra{\phi}\tilde{G}'^{\pm}_{\mp\frac{1}{2}}\ket{\chi}_{\Phi}$ give the supercharge action to first order in perturbation theory and are shorthand for the three-point functions, see e.g.~\cite{Gaberdiel:2023lco},
\begin{align}\label{eq:regulated_transition}
\bra{\phi}\tilde{G}'^{\pm}_{\mp\frac{1}{2}}\ket{\chi}_{\Phi} :=\pi\,\delta_{h_\phi,h_\chi}\, \corr{
\phi^\dagger(\infty)\,
V\bbr{\tilde{G}'^{\pm}_{\mp\frac{1}{2}}\Phi,1}\, \chi(0)
}\ ,
\end{align}
where $\delta_{h_\phi,h_\chi}$ requires that $\ket{\phi}$ and $\ket{\chi}$ have the same left-moving conformal dimension. All three operators in this correlator lie in a twisted sector.

\subsection{Covering surface}

To compute the transition in Eq.~(\ref{eq:regulated_transition}), we have to calculate a correlator
involving the single-cycle $w$-twisted field $\phi^\dagger$, the field $\tilde{G}'^{\pm}_{\mp\frac{1}{2}}\Phi$ from the $2$-cycle twisted sector, and the field $\chi$ in the $(w-k,k)$  multi-cycle twisted sector. This can be achieved using the covering map method of \cite{Lunin:2000yv}, where the correlator is lifted to a branched cover of the sphere on which the twist is implemented geometrically.

For the situation at hand, the covering surface is a sphere and the covering map is given by the polynomial
\begin{equation}\label{eq:cov_map}
\Gamma_{w,k}(z) = \br{\frac{k-w}{k}}^k\,z^{w-k}\,\br{z-\frac{w}{w-k}}^k\ .
\end{equation}
This map is branched over $0$, $1$, $\infty$, corresponding to the $\chi$, $\tilde{G}'^{\pm}_{\mp\frac{1}{2}}\Phi$, and $\phi^\dagger$ insertions, respectively. The points $z=0$ and $z=\frac{w}{w-k}$ are ramified preimages of $0$, with ramification indices $w-k$ and $k$, implementing the multi-cycle twist geometrically. The covering surface is sketched in Figure~\ref{fig:cov_surf}.

\subsection{Lifting of correlators}

Using the covering surface, the transition amplitudes (\ref{eq:regulated_transition}) can be written as correlators on the covering surface,
\begin{align}\label{eq:lifted_transition}
\bra{\phi}&\tilde{G}'^\pm_{\mp\frac{1}{2}}\ket{\chi}_\Phi = \\
&= \pi\,\delta_{h_\phi,h_\chi}\, \corr{\hat{\phi}^\dagger(\infty)\hat{\chi}_{k}(\tfrac{w}{w-k})\, \bbr{\widehat{\tilde{G}'^\pm_{\mp\frac{1}{2}}\Phi}}(1) \hat{\chi}_{w-k}(0)}_\text{cov}\ , \nonumber
\end{align}
where hats denote the lifts of the fields via the covering map $\Gamma_{w,k}$. Note that $\chi$ is lifted to \textit{two} states $\hat{\chi}_{w-k}$, $\hat{\chi}_k$ on the covering surface, as it lies in a multi-cycle twisted sector. Furthermore, the correlator is radially ordered.

In practice, one writes $\ket{\chi}$ as a fractional mode descendant of a primary operator, as in Eq.~(\ref{eq:example_states}). Each such mode is lifted by changing coordinates using $\Gamma_{w,k}$. Let us for example consider the action of a bosonic fractional mode $\ket{\chi}=\alpha_{-q}\ket{\sigma}$. If $q\in\frac{1}{k}\mbb{Z}$, i.e.~if the mode acts on the $k$ cycle, the element becomes a contour-integrated correlator
\begin{align}
\bra{\phi}&\tilde{G}'^\pm_{\mp\frac{1}{2}}\alpha_{-\frac{n}{k}}\ket{\sigma}_\Phi =\nonumber\\
&= \cint{C(\frac{w}{w-k})} dz\,
\br{\Gamma_{w,k}(z)}^{-\frac{n}{k}}\,\pi\,\delta_{h_\phi,h_\chi}\\
&\qquad \times\corr{\hat{\phi}^\dagger(\infty)\hat{\alpha}(z)\hat{\sigma}_{k}(\tfrac{w}{w-k})\, \bbr{\widehat{\tilde{G}'^\pm_{\mp\frac{1}{2}}\Phi}}(1) \hat{\sigma}_{w-k}(0)}_\text{cov}\ . \nonumber
\end{align}
where $\hat{\alpha}$ is the free boson on the covering surface. If the mode acts on the $w-k$ cycle instead, i.e.~$q\in\frac{1}{w-k}\mbb{Z}$, the contour integral over the position $z$ of $\hat{\alpha}$ instead runs around $z=0$.

Iterating this procedure, one can write the elements $\bra{\phi}\tilde{G}'^\pm_{\mp\frac{1}{2}}\ket{\chi}_\Phi$ as contour integrals over simple correlators---free fields and BPS ground states. This calculation can then be automated to evaluate the large number of contractions necessary.

\fi

\begin{center}
\large\textbf{Appendix: Statistical Analysis}
\end{center}

In this appendix we mention some technical details about the unfolding of the eigenvalues and provide additional statistical analysis of the level spacings.

\section{Unfolding}

To unfold the non-planar corrections to the spectrum, we choose a special prescription, which we now describe.
The purpose of unfolding the spectrum is to remove smooth mean density variations while retaining local spectral correlations.
This separation is only sensible if the mean density varies much slower than the local spectral fluctuation, which is typically not the case near the edges of the spectrum. Usually, a fixed number of the edge eigenstates is removed to deal with this issue. However, this procedure is subtle for the non-planar corrections, as they are obtained from an ensemble of matrices $\tilde{\gamma}_1|_{\mcl{E}}$ of different sizes. We adopt a prescription where, for each eigenspace, we only keep the portion of states whose mean density variation lies below 10 \%. We have checked the accuracy of this prescription by comparing the average gap ratio $\langle r \rangle$ before and after unfolding, which is accurate to a difference of less than $10^{-4}$.

\section{Gap ratio distribution and Cramer--von Mises distance}
To give a more fine-grained test of the integrable and RMT behaviour of the spectra, we have also calculated the distributions of gap ratios $r$ , and compared them to the known results for Poisson statistics and GOE \cite{Atas:2013gvn}. More specifically, we have calculated the Cram\'er--von Mises (CvM) distance (using the {\tt Mathematica} function {\tt CramerVonMisesTest} with option {\tt TestStatistic}) between our empirical level spacing (ls) and gap ratio (gr) distributions on the one hand, and  those describing Poisson statistics and GOE on the other. The results are summarized in Table~\ref{tab:CvM}: the CvM distance to the Poisson distribution is small (0.11 - 0.64) for the planar spectra, both for the fermions and the bosons. For the non-planar corrections to the fermionic spectrum, the CvM distance to GOE is quite small (0.27 and 0.39), while the situation for the bosonic spectrum is less definite. This supports our claim  that the planar spectra obey Poisson behaviour for both bosons and fermions, while the non-planar spectra of the fermions are close to GOE, with the bosons lying in a crossover regime.

{
\begin{center}
\begin{table}[h!]
\begin{tabular}{l | c | c | c | c}
 & Poisson, ls & GOE, ls & Poisson, gr & GOE, gr \\
\hline
bosons, planar & 0.18 & 42 & 0.11 & 64 \\
bosons, non-planar & 47 & 9.5 & 19 & 36 \\
fermions, planar & 0.12 & 22 & 0.64 & 44 \\
fermions, non-planar & 16 & 0.27 & 18 & 0.39
\end{tabular}
\caption{The CvM distances between the empirical level spacing (ls) and gap ratio (gr) distributions and the corresponding surmises of systems with Poisson and GOE statistics \cite{Haake:2010fgh,Atas:2013gvn}.}\label{tab:CvM}
\end{table}
\end{center}
}

\end{document}